\documentclass[iop]{emulateapj}
\usepackage{color}

\begin{document}
\newcommand{\ditto}[1][.4pt]{\xrfill{#1}~\textquotedbl~\xrfill{#1}}
\newcommand{\kms}{\mbox{km~s$^{-1}$}}
\newcommand{\s}{\mbox{$''$}}
\newcommand{\mloss}{\mbox{$\dot{M}$}}
\newcommand{\my}{\mbox{$M_{\odot}$~yr$^{-1}$}}
\newcommand{\ls}{\mbox{$L_{\odot}$}}
\newcommand{\ms}{\mbox{$M_{\odot}$}}
\newcommand\mdot{$\dot{M}  $}
\newcommand{\codos}{$^{12}$CO}
\newcommand{\doce}{$^{12}$CO}
\newcommand{\cotres}{$^{13}$CO}
\newcommand{\keunit}{ergs}
\newcommand{\momunit}{\ms\,\kms}
\newcommand\T{\rule{0pt}{2.6ex}}       
\newcommand\B{\rule[-1.2ex]{0pt}{0pt}} 
\title{A Molecular-Line Study of the Interstellar Bullet Engine IRAS05506+2414}
 
\author{Raghvendra Sahai\altaffilmark{1}, Chin-Fei Lee\altaffilmark{2}, Carmen S{\'a}nchez
Contreras\altaffilmark{3}, Nimesh Patel\altaffilmark{4}, Mark R. Morris\altaffilmark{5}, and Mark 
Claussen\altaffilmark{6}
}
\altaffiltext{1}{Jet Propulsion Laboratory, MS\,183-900, Caltech, Pasadena, CA 91109}
\altaffiltext{2}{Academia Sinica Institute of Astronomy and Astrophysics, P.O.  Box 23-141, Taipei 106, Taiwan}
\altaffiltext{3}{Dpto. de Astrof\'{\i}sica Molecular e Infraroja, Instituto de Estructura de la Materia-CSIC, Serrano 121, 28006 Madrid, Spain}
\altaffiltext{4}{Harvard-Smithsonian Center for Astrophysics, Cambridge}
\altaffiltext{5}{Division of Astronomy \& Astrophysics, UCLA, Los Angeles, CA 90095-1547}
\altaffiltext{6}{National Radio Astronomy Observatory, 1003 Lopezville Road, Socorro, NM 87801}

\email{raghvendra.sahai@jpl.nasa.gov}
 
\begin{abstract}
We present interferometric and single-dish molecular line observations of the interstellar 
bullet-outflow source IRAS05506+2414, whose wide-angle bullet spray is similar to the Orion BN/KL 
explosive outflow and likely arises from an entirely different mechanism than the classical 
accretion-disk-driven bipolar flows in young stellar objects. The bullet-outflow source is 
associated with a large pseudo-disk and three molecular outflows -- a high-velocity outflow (HVO), a 
medium-velocity outflow (MVO), and a slow, extended outflow (SEO). The size (mass) of 
the pseudo-disk is 10,350\,AU$\times$6,400\,AU (0.64--0.17\,\ms); 
from a model-fit assuming infall and rotation we derive a central stellar mass of 8--19\,\ms. The 
HVO (MVO) has an angular size $\sim5180$ ($\sim3330$) AU, and a projected outflow 
velocity of $\sim140$\,\kms~($\sim30$\,\kms). The SEO size (outflow speed) is $\sim 0.9$\,pc 
($\sim6$\,\kms). The HVO's axis is aligned with (orthogonal to)
that of the SEO (pseudo-disk). The velocity structure of 
the MVO is unresolved. The scalar momenta in the HVO and SEO are very similar, suggesting that the 
SEO has resulted from the HVO interacting with ambient cloud material. The  
bullet spray shares a common axis with the pseudo-disk, and has an age comparable to that of MVO 
(few hundred years), suggesting that these three structures are intimately linked together. We 
discuss several models for the outflows in IRAS\,05506+2414 (including dynamical decay of a stellar 
cluster, chance encounter of a runaway star with a dense cloud, and close passage of two 
protostars), 
and conclude that 2nd-epoch imaging to derive proper motions of the bullets and nearby stars can 
help to discriminate between them.
\end{abstract}


\section{Introduction}
High-mass ($M>8$\,\ms) stars play a decisive role in the evolution of
galaxies; throughout their life-cycles, they inject large amounts of energy and momentum into their
environments through massive stellar winds, UV radiation, and finally in their deaths as supernovae (Zinnecker and Yorke 2007). In
the earliest, very poorly understood, stages of their lives, massive stars and their immediate environments are
characterised by high
luminosities ($\gtrsim 10^4$\ls), dense and warm molecular gas, strong far-infrared dust emission (e.g.,
Sridharan et al. 2002), and very weak or no free-free continuum emission at cm wavelengths
(as they have not yet developed an ultracompact HII region).

Like their low-mass counterparts, high-mass protostars also exhibit bipolar outflows having varying degrees of
collimation and outflow luminosities (Qiu et al. 2008). An exciting recent result in the understanding of this
early evolutionary stage, based on a CO\,J=2-1 interferometric mapping study of the Orion BN/KL region (Zapata et al. 2009),
is that its enigmatic wide-angle outflow has been produced by a violent explosion during the dynamical decay of a
non-hierarchical massive
young stellar system $\sim$500\,yr ago (e.g., Bally et al. 2011: BCM11, Bally et al. 2017). A competing hypothesis is 
the chance encounter of a runaway star (BN) with
the KL star-forming core (Tan 2004, Chatterjee \& Tan 2012).

These are entirely different mechanisms from the classical accretion-disk-driven bipolar flows in YSOs, and begs the important question:
is the Orion BN/KL explosive outflow and its engine unique, or simply the most prominent example of a relatively commonplace phenomenon?  
The latter scenario appears more likely, given that most massive stars form in dense clusters where dynamic interactions between the most massive 
members may be common, and that 30\% of O-type stars are high-velocity runaways.

Our discovery, using HST, of the second clear-cut example of this phenomenon in our Galaxy, IRAS\,05506+2414 (hereafter IRAS\,05506)
(Sahai et al.\,2008\,[Setal08]), now provides 
a new opportunity to test these models. A few additional examples of the explosive outflow phenomenon 
are listed in BCM11 (e.g., DR21, Zapata et al. 2013), but 
they are somewhat less compelling, and/or lack the detailed similarity with Orion BN/KL. 

The HST optical and near-infrared images of IRAS\,05506 show a bright, compact central source (Sa: Setal08)
with a jet-like extension (Sa-jet: Setal08), and a fan-like spray of elongated knots which appear to emanate from it. Optical spectra show 
that these knots are moving at high velocities (radial velocity 350\,\kms) (Setal08). These structures are 
analogous to the near-IR ``bullets" seen towards the BN/KL source in the Orion nebula. Interferometric
observations of IRAS\,05506 at 2.6\,mm showed the presence of a high-velocity CO outflow (hereafter HVO) aligned with the optical jet
structure and roughly orthogonal to the bullet-spray. Setal08 also found extended NH$_3$ (1,1) emission
towards IRAS\,05506, which together with the combined presence of far-IR, H$_2$O and
OH maser, CO and CS J=2--1, emission, strongly suggested a dense, dusty star-forming core
associated with IRAS\,05506. Setal08 concluded that IRAS\,05506 is probably an intermediate-mass or massive
protostar. The lack of radio continuum and the
late G -- early K spectral type derived by Setal08 from the optical spectrum of this object (S{\'a}nchez Contreras et al. 2008) implied that this
star is very young since its temperature is still
too low to provide sufficient UV flux for ionisation. The 
time-scale of its bullet-outflow is very short (a few hundred years). 

In this paper, we present a new study of IRAS\,05506, based on 
extensive millimeter-wave observations of molecular lines, 
using both single-dish and high-angular resolution interferometric data. 
The plan of the paper is as follows. We first describe our observational results for the large-scale (arminute) cloud 
structure in which 
IRAS\,05506 is embedded, as seen in molecular-line and thermal dust emission observed with 
single apertures (\S\,\ref{lrgscal}). We then describe observational results for the small-scale (arcsecond) 
structure around the central source in IRAS\,05506 (\S\,\ref{smlscal}). Next, in 
\S\,\ref{struct}, we discuss the properties of the main physical components associated 
with IRAS\,05506, as derived from the above datasets -- these include a pseudo-disk and several outflows. 
In \S\,\ref{discus}, we compare IRAS\,05506 with the well-studied Orion 
BN/KL bullet-spray source, in the context of the published theoretical models that have 
been used to explain such sources. We present our conclusions in \S\,\ref{conclude}.

\section{Observations}
\subsection{SMA}
We obtained observations at $\lambda\sim1.3$\,mm with the SMA\footnote{The Submillimeter Array is a joint project
between the Smithsonian Astrophysical Observatory and the Academia Sinica
Institute of Astronomy and Astrophysics and is funded by the Smithsonian
Institution and the Academia Sinica} using the subcompact and extended array 
configurations. 
The subcompact track was obtained as ``filler" (i.e., the observations were carried out 
sharing the track with another project), with  phase center coordinates (J2000 RA,Dec): 
05:53:43.539, +24:14:45.2, using 7 antennas, with system temperatures, 
$T_{sys}\sim150-200$\,K. We shifted the phase center to 05:53:43.549, +24:14:43.999 in 
order to be consistent with the later extended tracks. The frequency coverage was 
$217-221$\,GHz in the LSB, and $229-233$\,GHz in the USB. Additional observational details (beam-size, orientation) are provided in 
Table\,\ref{tab:obs_log}. 

The doppler correction for LO frequency was based on an extragalactic source which was being observed for the main track, with which we shared our 
observations. This led to a shift of $+55$\,\kms~in the central velocity for our source (IRAS\,05506). Our visibility data were post-corrected using the task 
uti\_doppler\_fix in MIR-IDL\footnote{https://www.cfa.harvard.edu/~cqi/mircook.html} 
for this velocity shift.


With the extended array, data were taken over two tracks, one on 2012 January 20 (with 6 antennas, on-source time $\sim$6.5\,hr) and another on 2012 February 9 
(with 7 antennas, on-source time $\sim$8\,hr). During the extended array observations in Feb 2012, the atmospheric conditions were very good (atmospheric 
optical depth at 225\,GHz was around 0.03), and phase stability was also excellent. Since the system temperatures during January 20 ($T_{sys}\sim200-350$\,K) 
were much higher than in the February 9 ($T_{sys}\sim130-240$\,K), we have not used the former. 

Comparing the visibility amplitudes in the subcompact-array (hereafter ``subcom") data with the extended-array (hereafter ``ext") data, over similar u-v spacings, 
we find the extended amplitudes to be weaker by a factor of 0.85. Hence the visibilities in the ext data were scaled upwards by a 
factor 1/0.85 in order to make these datasets consistent, before 
combining them into a single dataset (hereafter ``extsub"), with high angular resolution ($\sim1\farcs4$) as well as sensitivity to extended
structures on a scale of $\sim10{''}$. 
In the rest of the paper, the default data cube used for presenting SMA observational results is the extsub one.


\subsection{OVRO}
We obtained observations at $\lambda\sim2.6$\,mm with OVRO using the H, E, and L array configurations; 
data were obtained with 2 tracks for each of arrays. 
The units of the digital spectral-line correlator were arranged to
provide a total bandwidth of 90\,MHz ($\sim$\,234\,\kms) with a
channel spacing of 1\,MHz (corresponding to $\sim$2.6\,\kms).  The
3\,mm continuum emission was observed simultaneously using the
dual-channel analog continuum correlator, which provided a total
bandwidth of 4\,GHz (after combining both IF bands).  Additional observational details (beam-size, orientation) are provided in 
Table\,\ref{tab:obs_log}. 

The calibration of the data was performed using the MMA software
package\footnote{MMA is written and maintained by the Caltech's
Millimeter Interferometry Group.}.  Data were gain-calibrated in
baseline-based mode using the quasar J0530$+$135, which was observed
at regular time intervals of $\sim$\,20\,minutes before and after our
target. The quasars 3C\,273 and 3C\,84 were used
as passband and flux calibrators. 

Reconstruction of the maps from the visibilities was done using the
Multichannel Image Reconstruction, Image Analysis and Display (MIRIAD)
software. We Fourier transformed the measured visibilities with robust
weighting (which is an optimized compromise between natural and
uniform weighting) for \doce\, and natural weighting for the continuum map for
S/N optimization. 


\subsection{Single-Dish Observations}\label{singledish}
We obtained single-dish observations of a variety of molecular lines (listed in Table\,\ref{tab:smt_obs_log}) towards IRAS\,05506 using the Arizona Radio 
Observatory's (ARO) SMT-10m and Kitt Peak 12m (KP12m) telescopes, and the Caltech Submillimeter Observatory's (CSO) 10-m 
telescope. These ``on-source'' spectra  
were obtained towards the coordinates of the optical/infrared source in IRAS\,05506 (i.e., Sa) using coordinates (J2000 RA, Dec): 5:53:43.6, 24:14:45, 
by 
position-switching the telescopes using an emission-free reference position. The beam-sizes (FWHM) for the SMT-10m at nominal frequences 
of 345 and 230\,GHz are $32{''}$ and $22{''}$,  respectively. For the KP12m (CSO), at 115 (461)\,GHz, the beam-size is $55{''}$ ($16\farcs5$). The 
beam-sizes scale with the wavelength over broad wavelength ranges within the 0.65, 0.8, 2.6 and 1.3 mm windows.

We mapped the $^{12}$CO and $^{13}$CO J=2-1 emission simultaneously towards a $\sim 11\farcm2\times5\farcm2$ area centered 
on IRAS\,05506 using the SMT's On-The-Fly (OTF) mapping observational mode (see, e.g., Bieging \& Peters 2011 for details of 
the technique). 

All data were calibrated using the standard chopper calibration method. 
Telescope pointing was checked frequently, i.e., every 45 to 60 minutes -- at the ARO, we used Mars as a continuum pointing source, and at the CSO, 
we used the 
circumstellar shell of the evolved star GX\,Mon as a line (CO J=2--1) pointing source. Additional relevant 
observing details (e.g., system temperatures, spectrometer backends, beam efficiencies) are given in Table\,\ref{tab:smt_obs_log}. 




\section{Results}
\subsection{Single-dish Data: The Large-Scale Environment of IRAS\,05506}\label{lrgscal}
\subsubsection{CO Observations}\label{smtco}
\begin{figure}[htbp]
\includegraphics[angle=0,origin=c,scale=0.7]{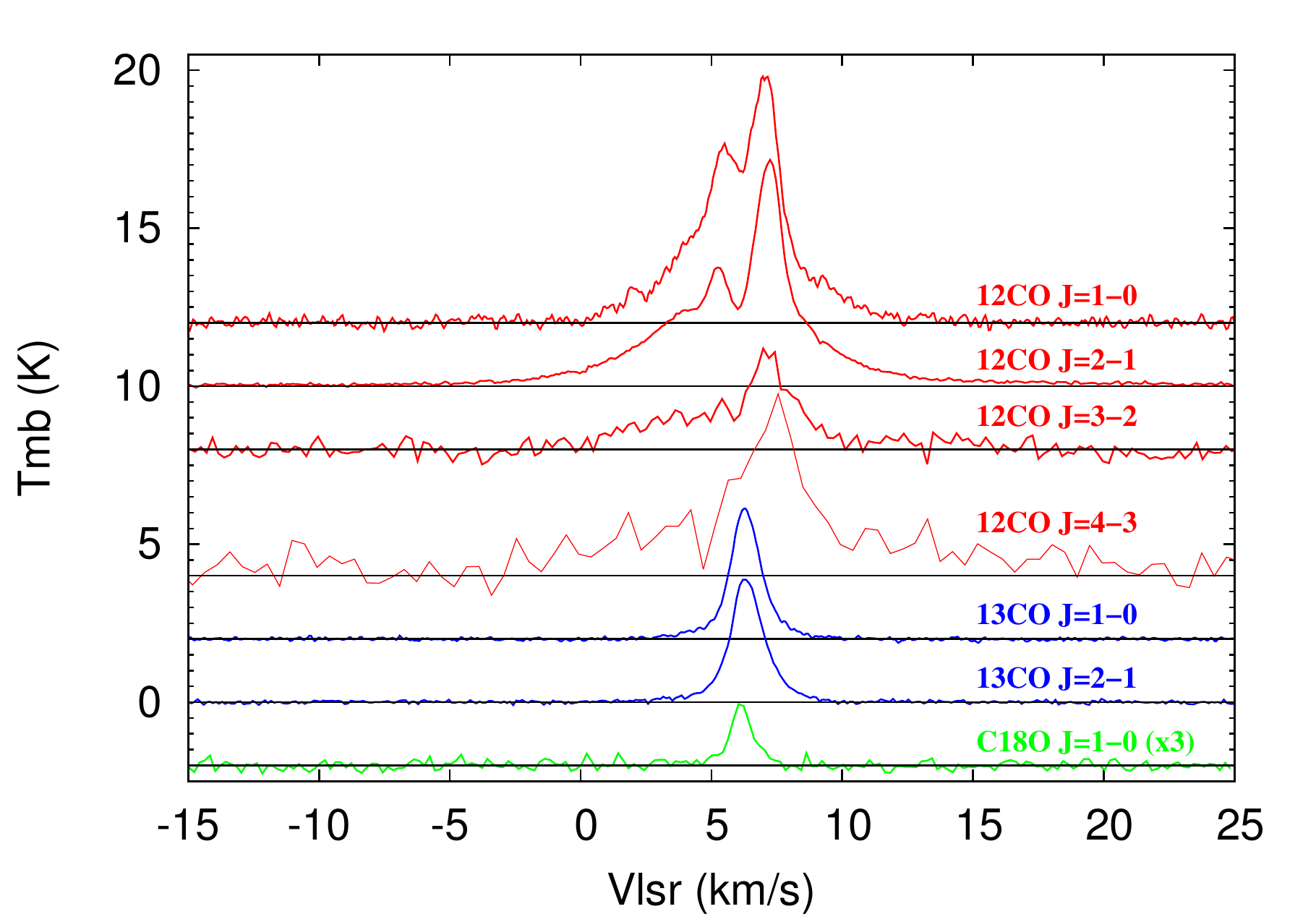}
\caption{Single-dish observations of IRAS\,05506 in the millimeter-wave lines from CO and its isotopologues, using the ARO's Kitt Peak 12-m and SMT and the CSO. 
Different transitions are shown in different colors and shifted vertically for clarity. The beam-sizes are different for different lines (\S\,\ref{singledish}).
The beam-sizes are different for different lines (\S\,\ref{singledish}).
} 
\label{co-13co-smt}
\end{figure}

The on-source $^{12}$CO J=2-1 and 1-0 lines show a double-peaked core, with a central minimum at $V_{lsr}\sim6$\,\kms, and prominent 
wings (Fig.\,\ref{co-13co-smt}). The on-source $^{12}$CO J=3-2 and 4-3 lines shows an emission peak near $V_{lsr}\sim7.4$\,\kms~with a FWHM velocity width of 
$\sim$2\,\kms, 
as well as a weaker, significantly broader component, having a FWHM of roughly 10 (15)\,\kms. In contrast, 
the corresponding $^{13}$CO 2-1 and 1-0 line profiles  
are centrally peaked at $V_{lsr}\sim6$\,\kms, with a FWHM=1.5\,\kms. 

\begin{figure}[htbp]
\hskip 0.4cm
\includegraphics[angle=0,origin=c,scale=0.7]{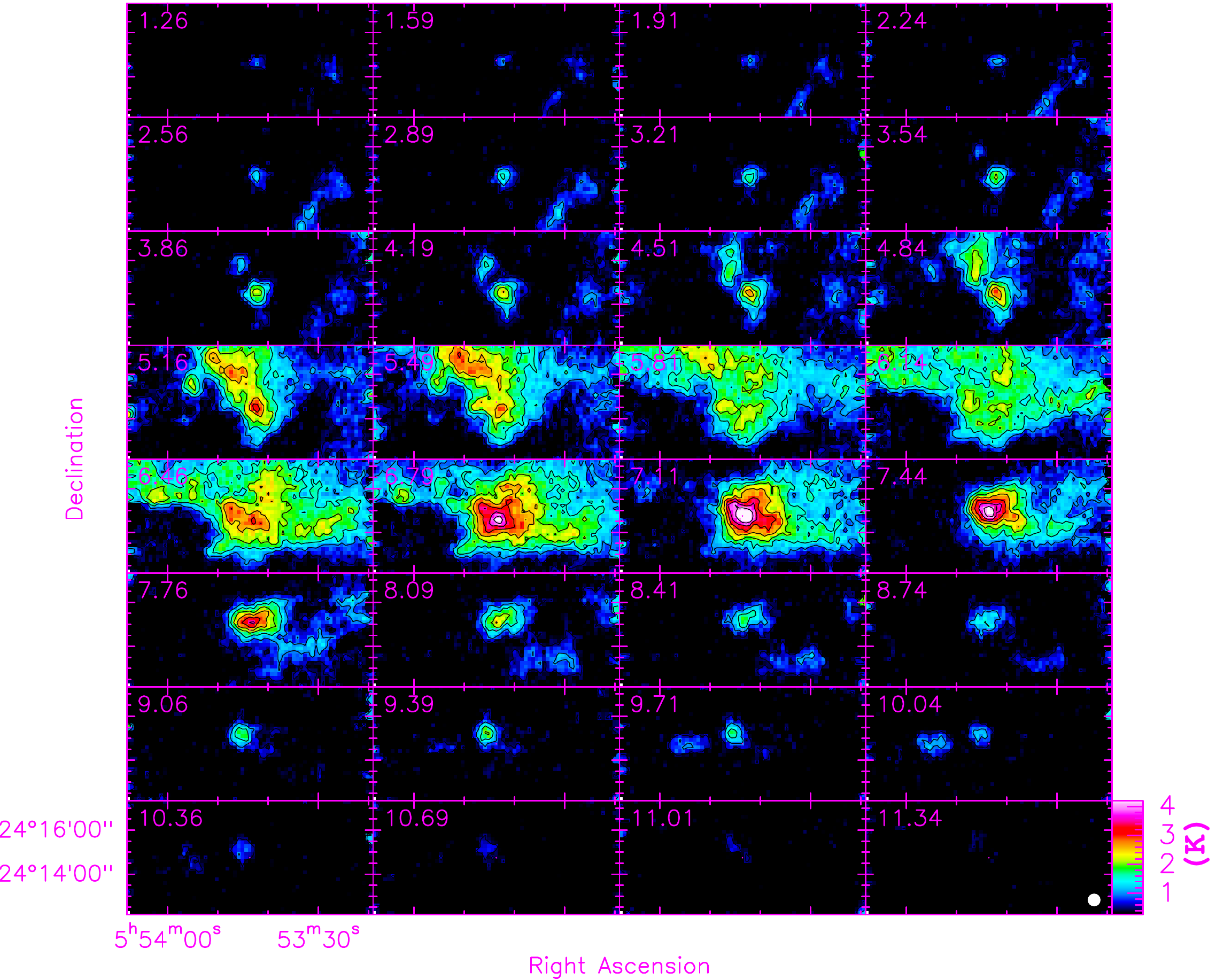}
\caption{SMT OTF channel map of $^{12}$CO J=2--1 emission from IRAS\,05506 and its large-scale environment, 
covering a $11\farcm2\times5\farcm2$ area. Each channel is 0.325\,\kms~wide, the beam (white circle in lower-right panel) 
is $32{''}$ (FWHM), contour levels are 0.5, 1.0, 1.5, 2.0, 2.5, 3.0 and 3.5\,K.
} 
\label{co21map-smt}
\end{figure}

\begin{figure}[htbp]
\hskip 0.4cm
\includegraphics[angle=0,origin=c,scale=0.7]{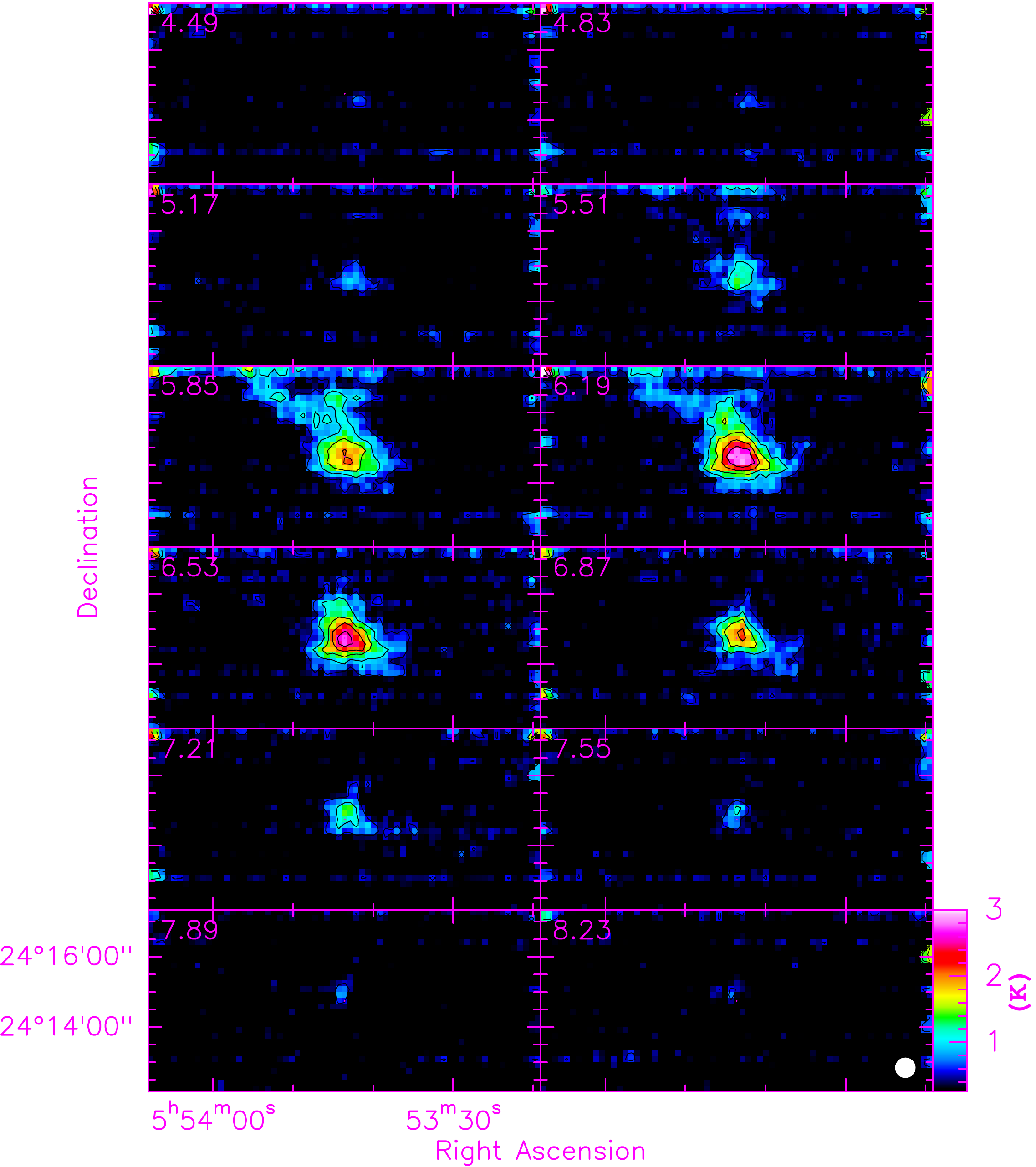}
\caption{SMT OTF channel map of $^{13}$CO J=2--1 emission from IRAS\,05506 and its large-scale environment, 
covering a $11\farcm2\times5\farcm2$ area (as in Fig.\,\ref{co21map-smt}). Each channel is 0.34\,\kms~wide, the beam (white circle in lower-right panel)  
is $33\farcs5$ (FWHM), and contour levels are 
0.5, 1.0, 1.5, 2.0 and 2.5\,K.
} 
\label{13co21map-smt}
\end{figure}

Our OTF mapping shows that $^{12}$CO J=2-1 emission is very extended, and at and around 
$V_{lsr}\sim6$\,\kms, it covers the full extent of 
the mapped area (i.e., $\sim 11\farcm2\times5\farcm2$)~at the $7.5\sigma$~noise level (1\,K intensity)  
(Fig.\,\ref{co21map-smt}). We will refer to this cloud as ``cloud A''; this cloud is likely a substructure of an even larger cloud, extending 
$30{'}$ EW and $10{'}$ NS, that was mapped by Setal08 in $^{12}$CO J=1-0 around IRAS\,05506 with ARO's KP12m dish.
The $^{13}$CO J=2-1 emission at $V_{lsr}\sim6$\kms~is much less extended (Fig.\,\ref{13co21map-smt}), and is 
strongly peaked towards IRAS\,05506 -- the contour at 30\% of the peak \cotres~intensity 
defines a roughly triangular-shaped region (with east-west and north-south aligned 
sides of length $\sim125{''}$). We will refer to this structure as ``cloud A core''. 

The on-source C$^{18}$O J=1-0 line profile has the narrowest width (FWHM$\sim$0.4\,\kms) amongst all the single-dish CO line profiles, is 
likely dominated by dense, cold gas in the cloud A core, whereas the other $^{12}$CO and $^{13}$CO J=1--0 line profiles likely include contributions from an 
outflow (see below). 

\begin{figure}[htbp]
\includegraphics[angle=0,origin=c,scale=0.7]{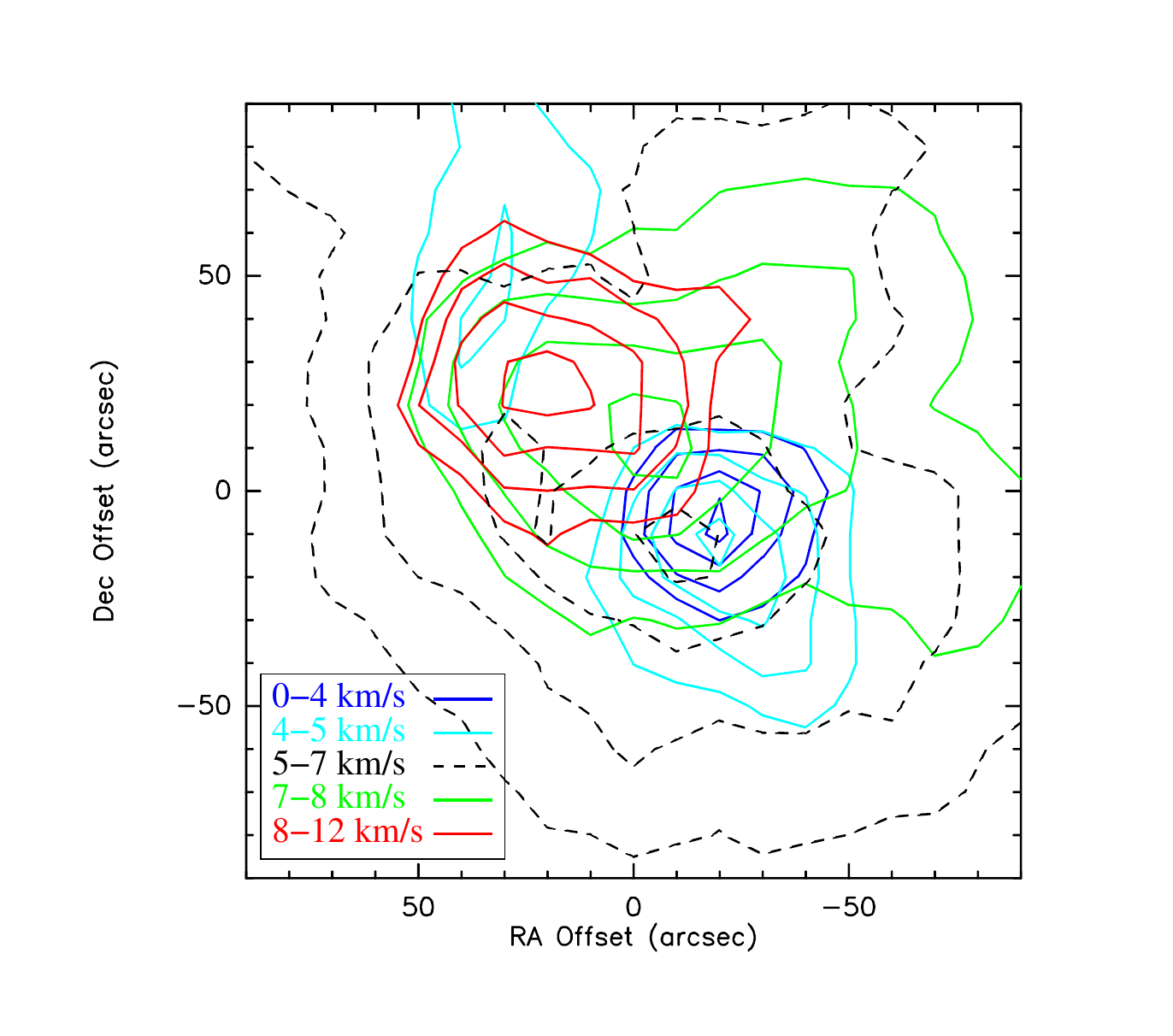}
\vskip -0.5cm
\caption{On-The-Fly (OTF) map of the $^{12}$CO J=2--1 emission towards IRAS\,05506, using the SMT 10-m, integrated over different velocity bins, showing the 
large-scale slow bipolar outflow (SEO). The beam size is $32{''}$. The map center is at (J2000 RA, Dec): 5:53:43.6, 24:14:45. 
The bins are $V_{lsr}=0-4$\,\kms~(blue), $4-5$\,\kms~(cyan), 
$5-7$\,\kms~(black), $7-8$\,\kms~(green), $8-12$\,\kms~(red), and the contour levels are at 0.5, 0.65, 0.8 and 0.95 of the peak integrated-intensity values 
for each velocity range (5.1, 2.7, 7.2, 4.0, 4.5 K--\kms, respectively).
} 
\label{co21-smt-blue-red}
\end{figure}

Using the OTF, we find that the emission in the blue wing 
($V_{lsr}=0-5$\,\kms) and red wing ($7-12$\,\kms), as seen in the on-source $^{12}$CO(2-1) profile, 
delineates two large extended bipolar outflow lobes around the location of IRAS\,05506, 
with the blue (red)-shifted lobe located SW (NE) of center (Fig.\,\ref{co21-smt-blue-red}). 
Hereafter, we refer to this outflow as the Slow Extended Outflow, or SEO. We do not find emission in our OTF map 
at velocities shortward (longwards) of $V_{lsr}\sim0$\,\kms ($V_{lsr}\sim12$\,\kms).

\subsubsection{Shock and High-Density Tracers}\label{txt:smtso}
\begin{figure}[htbp]
\includegraphics[angle=0,origin=c,scale=0.7]{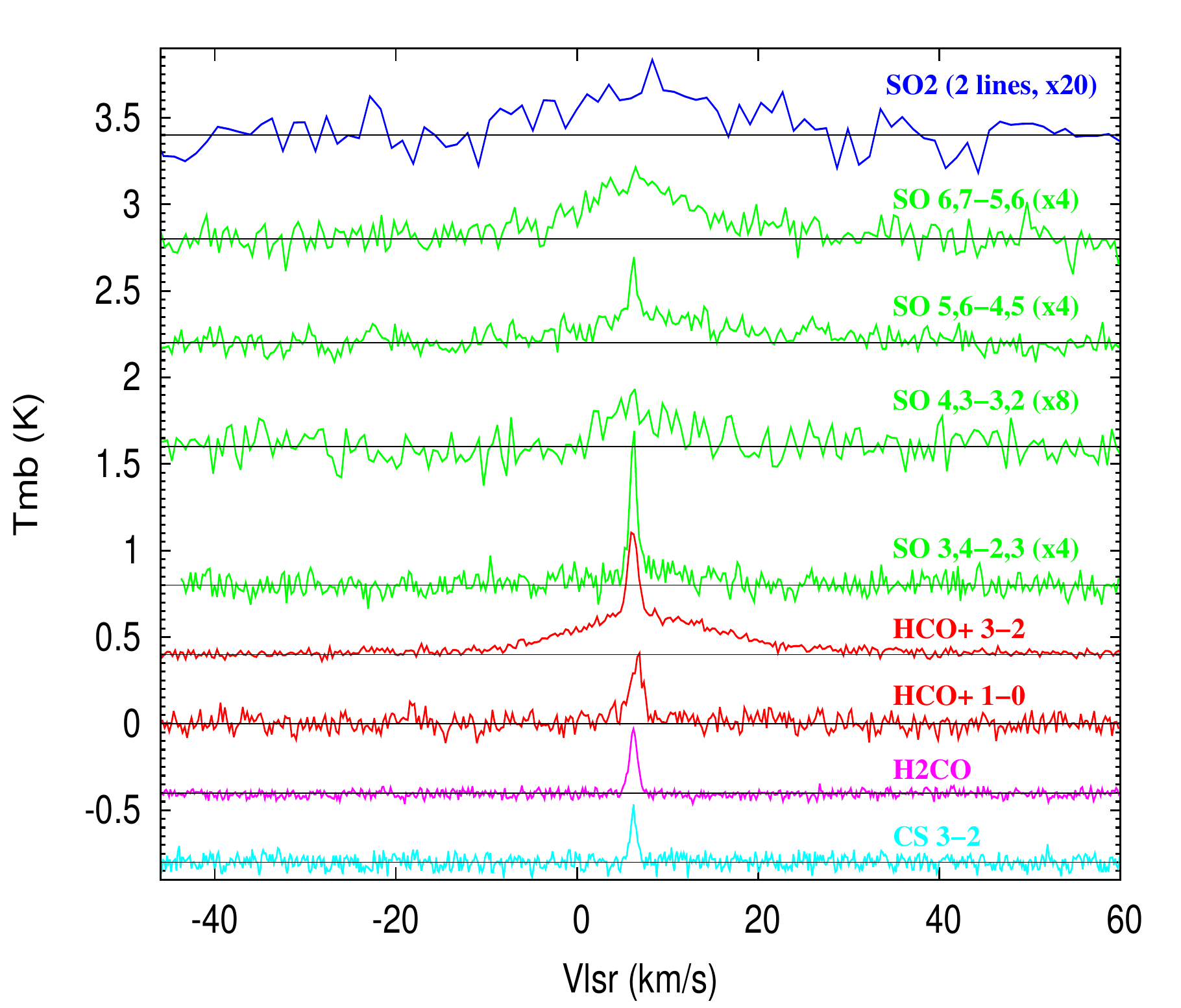}
\caption{Single-dish observations of IRAS\,05506 in the millimeter-wave lines that are probes of shocks and high-density gas, using the 
ARO's Kitt Peak 12-m and SMT 10-m telescopes. Different transitions are shown in different colors, scaled and shifted vertically for clarity.
The beam-sizes are different for different lines (\S\,\ref{singledish}).
} 
\label{fig:smt-12m-so-hco}
\end{figure}

Out of the four SO lines observed (N,J=6,7-5,6, N,J=5,6-4,5, N,J=4,3-3,2 and N,J=3,4-2,3: hereafter SO lines 1, 2, 3, \& 4, respectively), two (SO lines 2 and 
3) show a narrow central component, together with weak wings. 
The highest-excitation line, SO line 4, shows only the broad component, and the lowest-excitation line, SO line 1\footnote{the energies of the upper levels 
for SO lines 1--4 are respectively, 11.02, 19.93, 24.30, 33.04 cm$^{-1}$}, shows 
only the narrow component  (Fig.\,\ref{fig:smt-12m-so-hco}). 
We also detected weak, broad emission from the SO$_2$ J(K$_a$,K$_c$)=4(3,1)-4(2,2) and 12(1,12)-11(1,11) lines. 
The SO and SO$_2$ emission likely arises from shocked gas, as has been proposed in the case of the HH\,212 jet (Podio et al. 2015).

The HCO$^{+}$ J=3-2 line profile has a central narrow component with FWHM 1.3\,\kms, and a 
broad component with wings extending to $\sim-15$\,\kms~and $\sim30$\,\kms; the HCO$^{+}$ J=1-0 line shows only the narrow component.
We show later that the broad velocity wings of the SO lines 
likely arise in a very compact outflow in IRAS\,05506 (\S\,\ref{txt:sma-so}). The SO$_2$ and HCO$^{+}$ broad-wing emission which covers a similar velocity 
range as SO, most likely also comes from the same outflow.

The CS J=3-2 and H$_2$CO 2(1,1)-1(1,0) lines show only a narrow component centered at $V_{lsr}=6.3$\,\kms, which 
most likely comes from dense gas in the cloud A core.

\subsubsection{Continuum}\label{spire}
\begin{figure}[htbp]
\includegraphics[angle=0,origin=c,scale=0.9]{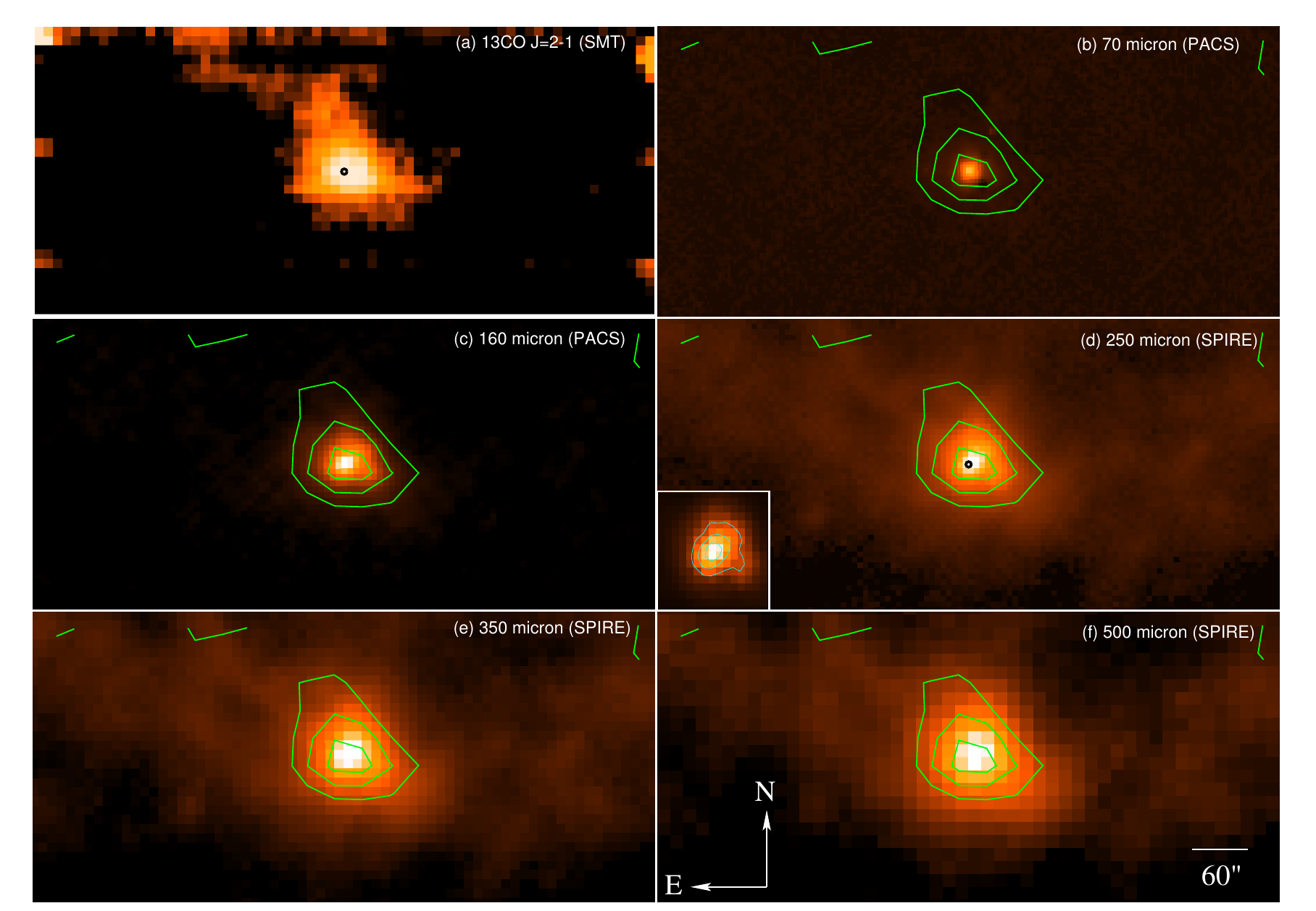}
\caption{IRAS\,05506 and its large-scale environment imaged with the SMT and Herschel, covering a $11\farcm2\times5\farcm2$ area (as in 
Fig.\,\ref{co21map-smt}). (a) SMT \cotres~J=2-1 map, 
(b,c) PACS 70 \& $160\,\micron$ image, (d,e,f) SPIRE 250, 350 
\& $500\,\micron$ images. The PACS and SPIRE images are overlaid with the \cotres~contours 
showing 75\%, 50\% and 30\% of the peak \cotres~J=2-1 intensity ($T_{A}^{*}=2.9$\,K). 
The location of the central source is marked by a black circle in the  \cotres~and $250\,\micron$ images. Inset in panel d shows a $1\farcm5\times1\farcm5$ 
region around the central source. A square-root stretch is used for the 
intensities in all panels except the inset, where a linear stretch is used. The angular-resolution (FWHM beam-size) in the Herschel images at 70, 160, 250, 
250, 
\& 500\,\micron, respectively, is 8.4, 13.5, 18.2, 24.9, and 36.3 arcsec, and the peak intensity of the source is 517.6, 126.5 43.6, 12.9 and 3.1 
mJy\,arcsec$^{-2}$.
}
\label{i05506herschel}
\end{figure}

IRAS\,05506 was observed with the SPIRE and PACS instruments onboard the Herschel Space Observatory (HSO) as part of the 
Hi-Gal survey (Molinari et al. 2016). We have downloaded the Level 2.5 images from the Herschel Science Archive. 
The SPIRE images (at 500, 350, and 250\,\micron) show a bright emission source roughly centered at the 
location of the 2MASS optical/infrared source in IRAS\,05506 (Sa) found by Setal08\footnote{J2000 RA, Dec: 05:53:43.56, 24:14:44.7} 
(Fig.\,\ref{i05506herschel}). The far-IR source has a morphology that is similar to that of the 
\cotres~J=2-1 emission from the cloud A core. Much fainter dust nebulosity can be seen extending roughly 
towards the ENE and W, with the same extent and morphology as the faint emission seen in 
the \codos~J=2-1 emission at the radial velocity of cloud A. 
The central peak appears to 
be elongated SE-NW (Fig.\,\ref{i05506herschel}d, inset). 
The half-power size of this peak is $31{''}\times21{''}$ (deconvolved size is 
$25{''}\times11\farcs5$, given the $17\farcs6$ SPIRE beam at 250\,\micron). 

The PACS images (70\,\micron~and 160\,\micron) show emission predominantly from a bright, round 
source with a half-power diameter of $10\farcs8$ at 70\,\micron, centered at the location 
of HST source Sa (deconvolved size is $9\farcs3$, given the $\sim5\farcs4$ PACS beam at 
70\,\micron). 
\subsection{Interferometric Data: The central source of IRAS\,05506}\label{smlscal}
We detected emission in the J=2-1 lines of $^{12}$CO, $^{13}$CO, and C$^{18}$O, and the SO N,J=5,6-4,5, and SiO J=5-4 (v=0) lines 
from IRAS\,05506 with the SMA. 
Additional, more extended tracks were observed towards IRAS\,05506 with the OVRO array\footnote{during the course of a survey of post-AGB objects by 
S{\'a}nchez Contreras \& Sahai 2012} than those presented in Setal08 for the $^{12}$CO and $^{13}$CO J=1-0 lines.

\subsubsection{CO Observations}\label{smaco}
\begin{figure}[htbp]
\includegraphics[angle=0,origin=c,scale=0.7]{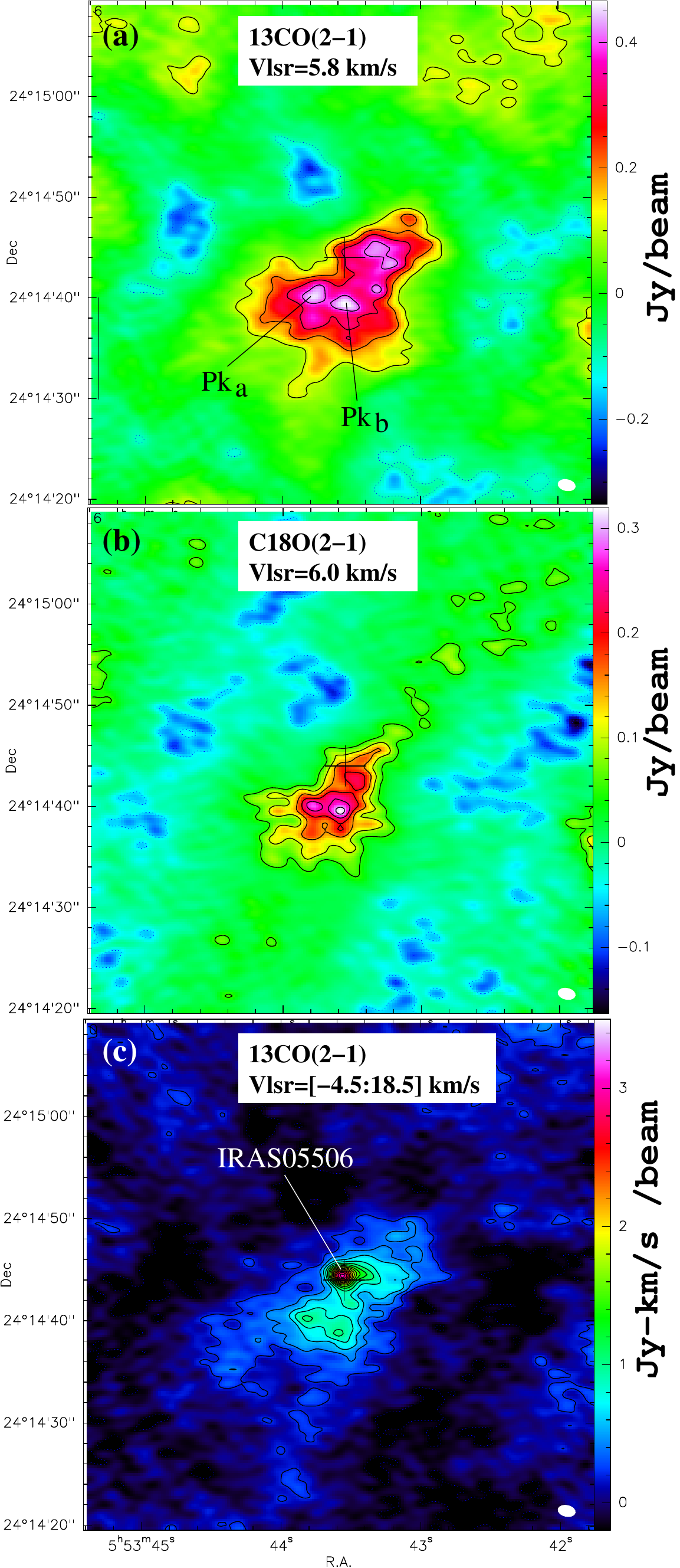}
\vskip 0.5cm
\caption{SMA map of IRAS\,05506 Cloud B (a) in $^{13}$CO J=2--1 emission at $V_{lsr}$=5.9\,\kms~(channel width is 1\,\kms); (b) 
C$^{18}$O J=2--1 emission at $V_{lsr}=6$\,\kms~(channel width is 1\,\kms); and (c) in $^{13}$CO J=2--1 emission 
in the range $V_{lsr}=-4.5$ to 18.5\,\kms. The beam FWHM is $1\farcs7\times1\farcs05$, $PA=78.1\arcdeg$ (white ellipse in lower-right corners of panels).
Scale bar shows intensity units in Jy\,beam$^{-1}$. The cross shows the location of the 
phase-center for the maps (J2000 RA, Dec) 05:53:43.55, 24:14:44.0.
}
\label{13co21cloud}
\end{figure}

IRAS\,05506 (Sa) is located near the north-east (NE) periphery of an extended, structured 
cloud of size $\sim15{''}\times7{''}$ -- the latter is best seen in the $^{13}$CO and C$^{18}$O J=2-1 maps, since these lines  
are more optically-thin than $^{12}$CO J=2-1 (Fig.\,\ref{13co21cloud}a,b). 
This cloud (hereafter ``cloud B'') has its long axis oriented roughly along the SE-NW direction as 
seen in the $^{13}$CO J=2-1 map at $V_{lsr}=6$\,\kms. 

We note that cloud B shows several distinct compact peaks in the $^{13}$CO and C$^{18}$O 
J=2-1 maps at $V_{lsr}=5-7$\,\kms, none of which are associated with the IRAS\,05506 Sa. 
These compact cores (the brightest of these are labelled $Pk_{a}$ and 
$Pk_{b}$) show no evidence for stars, as no associated point-sources are found in the 
near-IR (HST 0.8\,\micron~image, Setal08) or IR (2MASS J,H, K-band images, Setal08) or an 
archival Spitzer/IRAC 4.5\,\micron~image. In the 2MASS and Spitzer images, the wings of the PSF of the very bright 
IRAS\,05506 source at these wavelengths overlap the locations of these dense cores, making 
it difficult to discern faint counterparts. However, in the HST image this is not an 
issue, and althoough faint stars can be seen even closer to the bright IRAS\,05506 
source than $Pk_{a}$ and $Pk_{b}$ (Setal08) there are no optical counetrparts for these peaks. There is 
also no evidence for broad-velocity emission signifying the presence of outflows associated with these peaks.

\begin{figure}[htbp]
\includegraphics[angle=0,origin=c,scale=1.0]{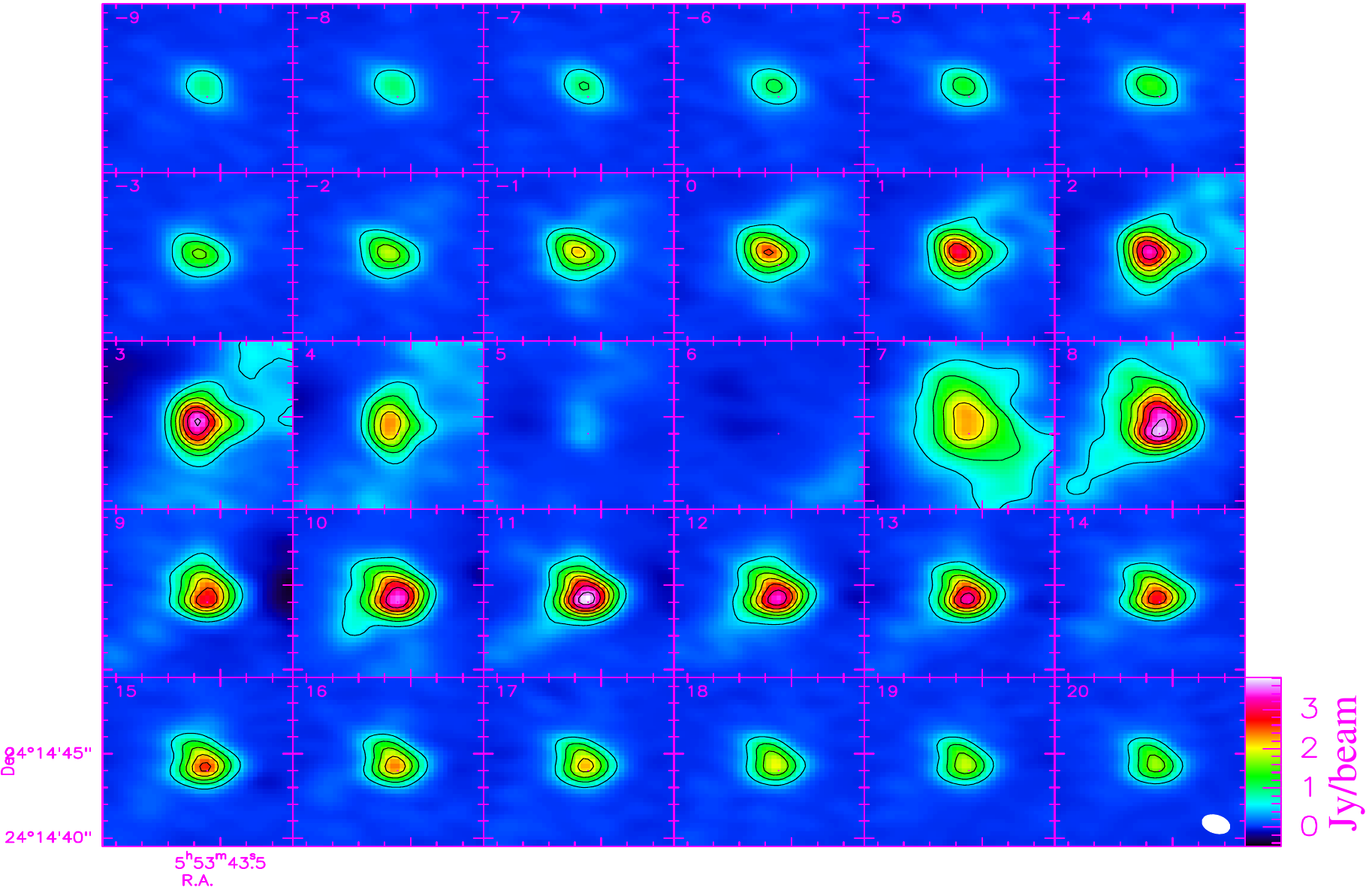}
\vskip 0.5cm
\caption{SMA $^{12}$CO J=2--1 channel map of IRAS\,05506 covering a $10{''}\times10{''}$ field-of-view and velocity range $V_{lsr}=-9$ to 20\,\kms.
The beam FWHM is $1\farcs6\times1\farcs0$, $PA=72.9\arcdeg$ (white ellipse in bottom-right panel).}
\label{sma-co21-chmap}
\end{figure}

At the location of IRAS\,05506, no emission peak is seen in the $^{13}$CO J=2-1 map at 
$V_{lsr}\sim6$\,\kms, but when the emission is integrated over a 
broad velocity range, V$_{lsr}=-4.5$ to 18.5\,\kms~(Fig.\,\ref{13co21cloud}c), a bright compact source becomes visible there. 
Channel maps of the $^{12}$CO J=2-1 emission, 
covering a central velocity range around IRAS\,05506's systemic velocity, i.e.,  $-9\le\,V_{lsr}(\kms)\le\,20$, show  
emission mostly confined to a rather compact structure, about $2\farcs5$ in size, centered at coordinates (J2000 RA, Dec): 
05:53:43.55, 24:14:45.1 
(Fig.\,\ref{sma-co21-chmap}), except in the $V_{lsr}$=5 and 6\,\kms~channels, where the emission is very weak or absent. This 
absence of emission results from strong absorption by cold gas in cloud A (described above, \S\,\ref{smtco}). 

\begin{figure}[htbp]
\includegraphics[angle=0,origin=c,scale=1.0]{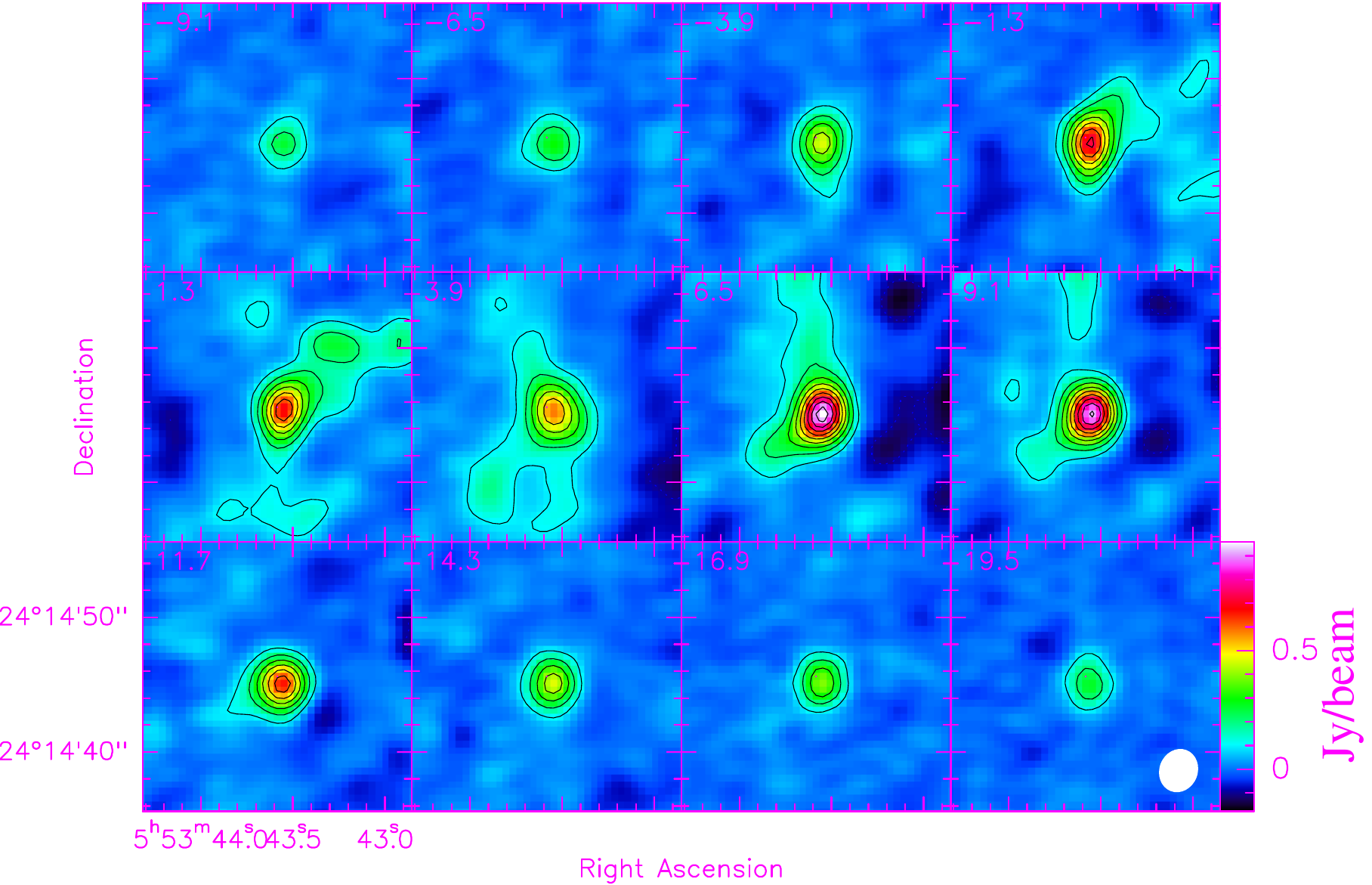}
\vskip 0.5cm
\caption{OVRO $^{12}$CO J=1--0 channel map of emission from IRAS\,05506  covering a $20{''}\times20{''}$ field-of-view 
and velocity range $V_{lsr}=-9.1$ to 29.9\,\kms. The beam FWHM is $3\farcs1\times2\farcs7$, $PA=-21\arcdeg$ (white ellipse in bottom-right panel).
} 
\label{ovro-co10-chmap}
\end{figure}

A similar compact source is seen in the OVRO $^{12}$CO J=1-0 map at the same location (Fig.\,\ref{ovro-co10-chmap}), 
but appears slightly more extended because of the larger beam size in these data. However, unlike J=2-1, the J=1-0 data show the compact 
source also in the $V_{lsr}=6.5$\,\kms~(2.6\,\kms~wide) channel, which overlaps the velocity range of the $V_{lsr}$=5 and 6\,\kms~channels in the 
SMA map. We infer that the foreground cloud A has a significantly lower optical depth in $^{12}$CO J=1-0, compared to J=2-1, along the  
line-of-sight towards the compact source.
The compact source in IRAS\,05506 is not seen in C$^{18}$O J=2-1 or $^{13}$CO J=1-0, either at the systemic velocity or when integrated over a 
broad-velocity range (as above, V$_{lsr}=-4.5$ to 18.5\,\kms), and we conclude that the emission from the source in the lines of 
CO and its isotopologues is dominated by relatively optically-thin emission extending over a broad velocity range.

\begin{figure}[htbp]
\includegraphics[angle=0,origin=c,scale=0.6]{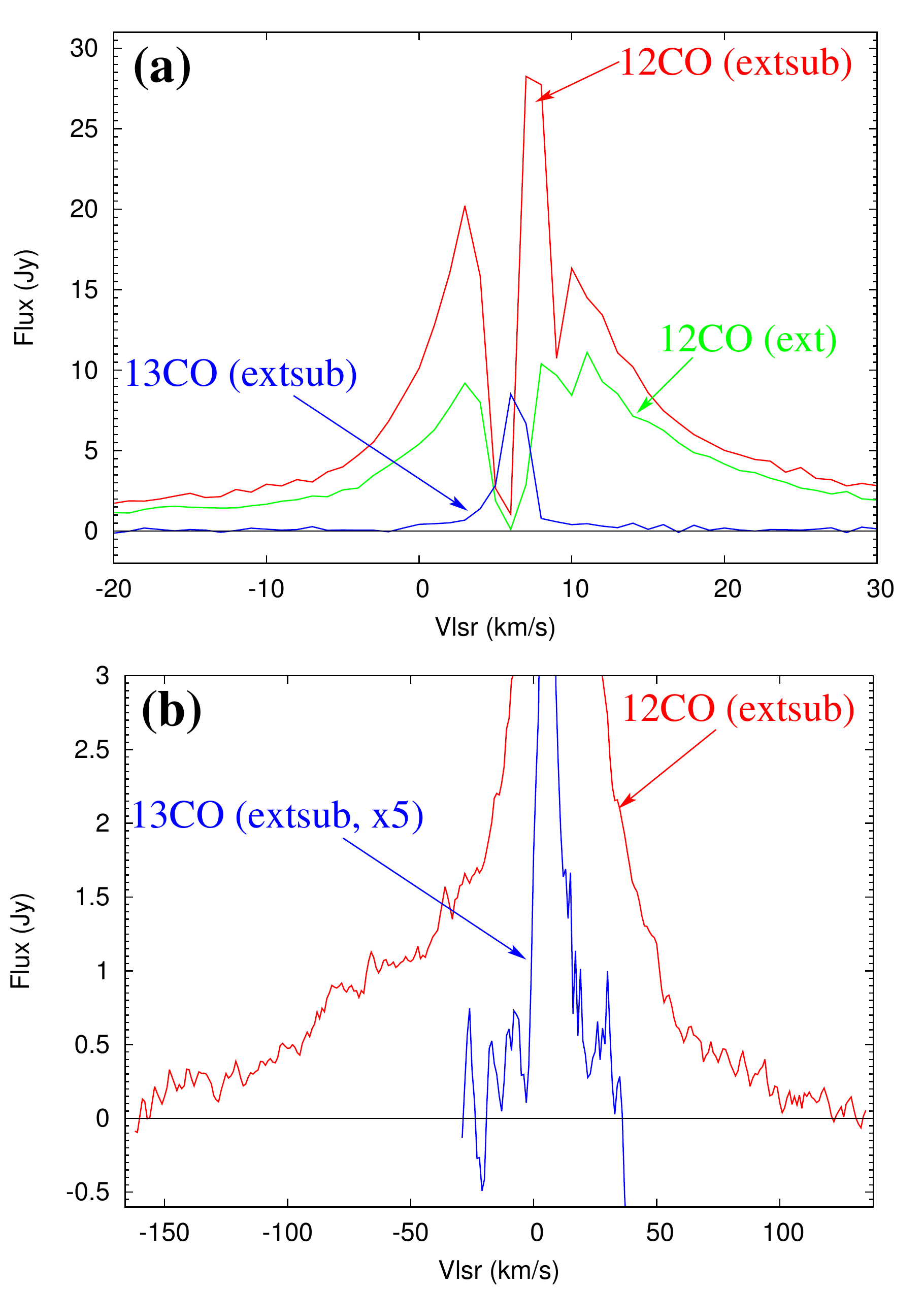}
\caption{(a) Spatially-integrated $^{12}$CO and $^{13}$CO J=2-1 spectra extracted from a $10{''} \times 10{''}$ box 
centered on IRAS\,05506, using the SMA data. For $^{12}$CO, the
profiles have been extracted from the extsub ({\it red}) and ext ({\it green}) 
datacubes, and for $^{13}$CO ({\it blue}), the
profile has been extracted from the extsub array datacube. 
(b) The $^{12}$CO and $^{13}$CO J=2-1 extsub profiles in panel $a$ (smoothed using a 3-point boxcar 
function), shown over a wider velocity range and a magnified intensity scale to 
show the faint high-velocity wings clearly (the $^{13}$CO spectrum has been scaled up by a factor 5).
}
\label{co13co-spec}
\end{figure}

In order to understand the kinematic structure of this compact CO source, 
we have extracted a $^{12}$CO J=2-1 spectrum from the extsub datacube 
using a $10{''} \times 10{''}$ box-aperture centered on IRAS\,05506. The spectrum shows a 
triple-peaked
structure in its core (Fig.\,\ref{co13co-spec}a), which is a combination of two features: the first is a broad absorption 
feature extending over $V_{lsr}\sim3-10$\,\kms~due to 
foreground absorption from cloud A (and/or cloud A core), and the second is emission in the $V_{lsr}\sim7-8$\,\kms~range from an extended structure 
evident in the 
\codos~channel map at these velocities (see Fig.\,\ref{sma-co21-chmap}) (hereafter ``cloudlet B1'').

The $^{13}$CO J=2-1 spectrum does not show the triple-peaked structure seen in $^{12}$CO J=2-1. It is centrally-peaked with 
a narrow core of FWHM 1.8\,\kms (1.5\,\kms~deconvolved), a broad component extending about $\pm$10\,\kms~from the peak at $V_{lsr}=6-7$\,\kms. 
The broad component comes from the compact source identified above in the \cotres~J=2-1 moment 0 map (Fig.\,\ref{13co21cloud}c). 

\begin{figure}[htbp]
\hskip -2in
\includegraphics[angle=-90,origin=c,scale=0.95]{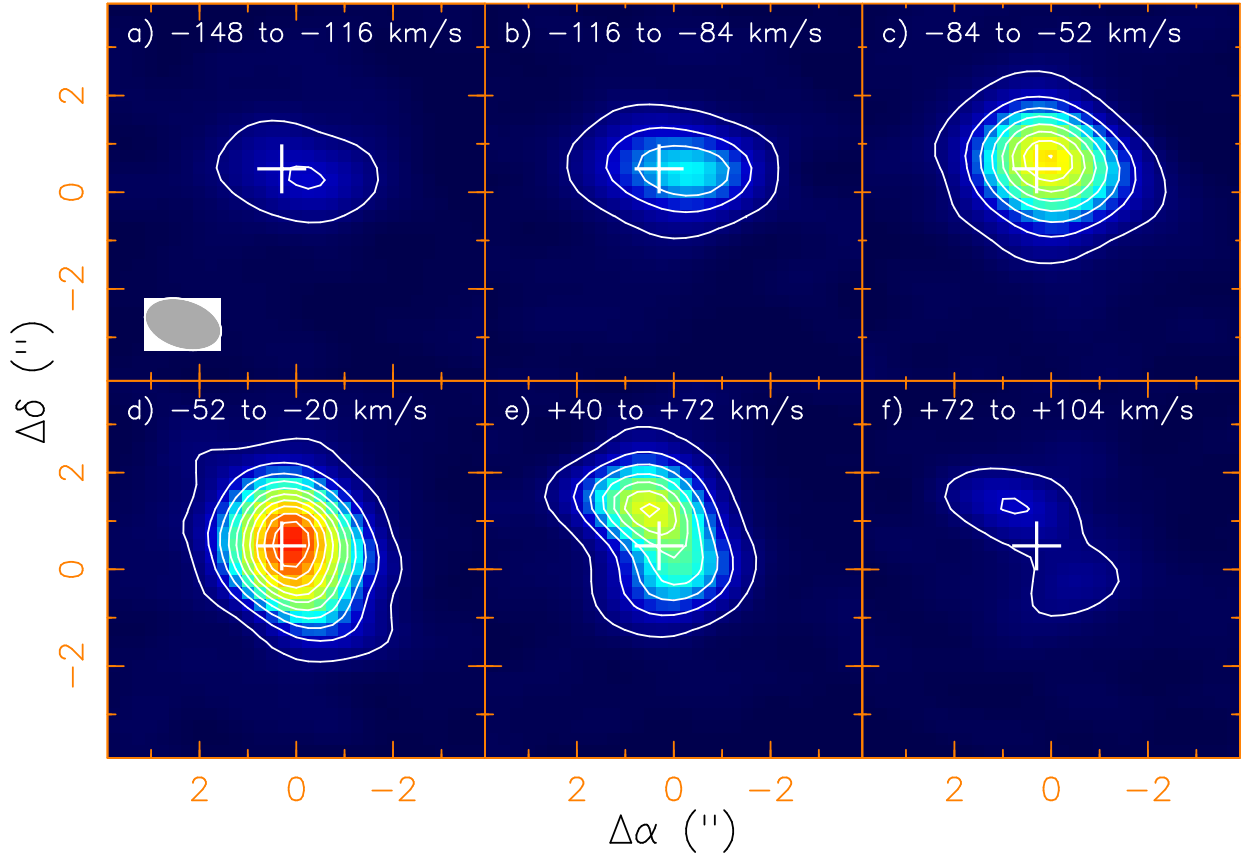}
\caption{SMA map of the $^{12}$CO J=2-1 emission from IRAS\,05506 integrated over the extreme blue-shifted and red-shifted velocity ranges. The cross shows the 
location of the optical/IR source Sa. Contour levels are $0.8+1.3\,n$\,Jy\,beam$^{-1}$\,\kms, with  n=0,1,2,... Beam is shown as grey ellipse in panel $a$.
}
\label{extremevel}
\end{figure}

The \codos~spectrum shows very extended wings seen out to $V_{lsr}$ of $\sim -155$\,\kms~in the blue wing 
and $\sim 120$\,\kms~in the red wing (Fig.\,\ref{co13co-spec}b)
In order to elucidate the structure of the faint emission in the line wings at $V_{lsr}<-20\,\kms$ and $V_{lsr}>40\,\kms$, we plot 
the integrated emission in selected velocity intervals covering the line wings (Fig.\,\ref{extremevel}).
The strongest blue-shifted and red-shifted emission $^{12}$CO J=2-1 emission at large velocity offsets from the systemic velocity (i.e., 
$V_{lsr}=-52$ to $-20$\,\kms~in the blue and $V_{lsr}=40$ to $72$\,\kms~in the red) is 
separated by $\sim0\farcs9$ along an axis that is roughly oriented along $PA\sim$40\arcdeg, with the blue-shifted emission in the south-west (SW). At more extreme velocity offsets (i.e., 
$V_{lsr}<-52$\,\kms~and $V_{lsr}>72$\,\kms) the orientation of the vector 
joining the blue- and red-shifted emission shifts to a larger $PA$, $\sim$55\arcdeg, but the uncertainty in this value is large because the red-shifted 
emission at these velocities 
is very weak, and because a weaker, secondary red-shifted peak is found located to 
the SW of the blue-shifted peak by $\sim1\farcs6$. We conclude that although the HVO direction in the SMA $^{12}$CO J=2-1 data is 
roughly consistent with that inferred by Setal08 
from their OVRO $^{12}$CO J=1-0 data, the HVO is likely more complex than a single bipolar outflow. 

\begin{figure}[htbp]
\includegraphics[angle=-90,origin=c,scale=0.7]{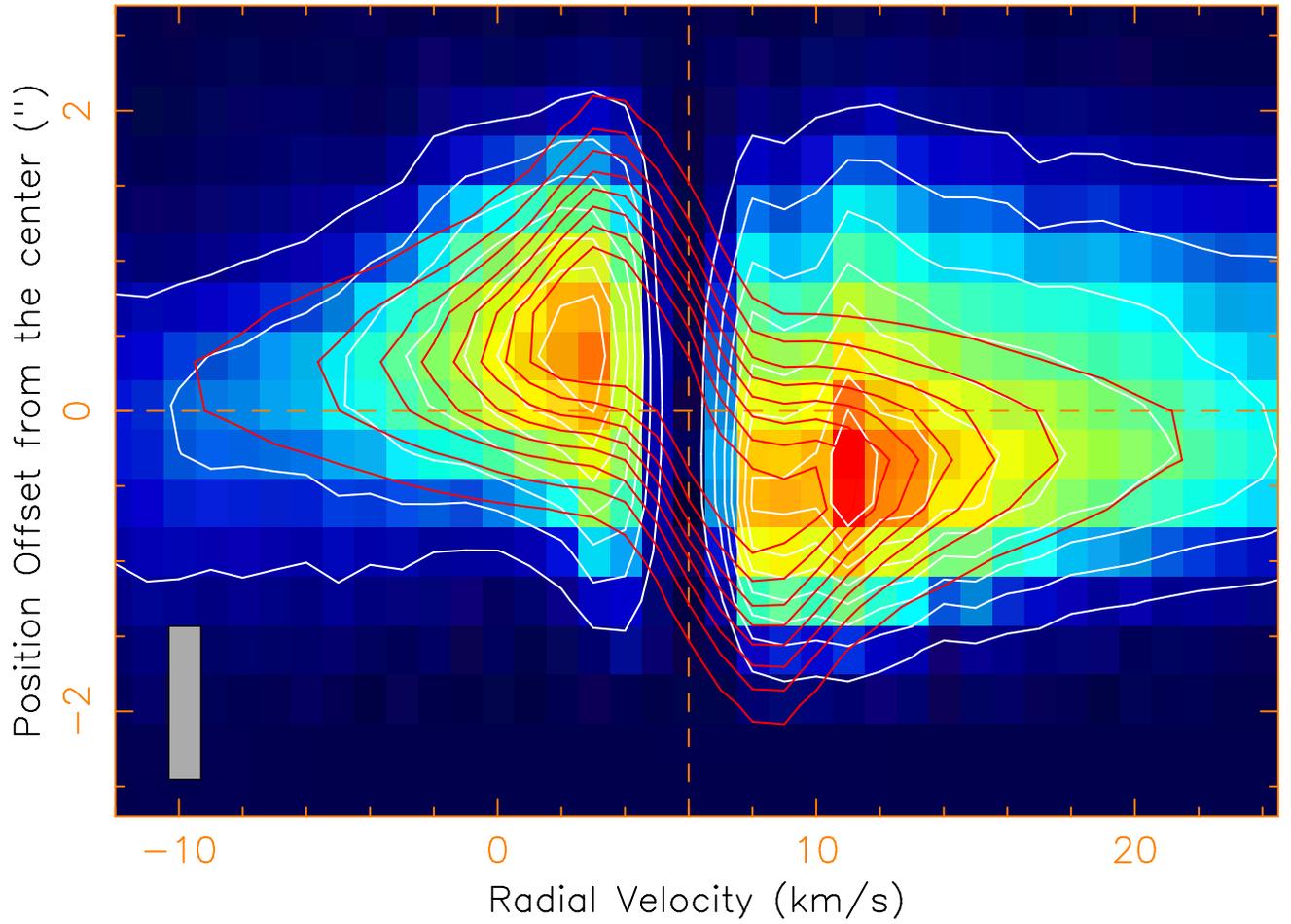}
\caption{Observed and model position-velocity plots of the \codos~intensity along a cut at $PA=40$\arcdeg. Colorscale and white 
contours, with levels 
$0.3\,n$\,Jy\,beam$^{-1}$, n=1,2,... , show the observed intensity extracted from the SMA $^{12}$CO J=2--1 ext datacube. Negative (positive) postion offsets 
are 
north-east (south-west) of center. 
The intensity from a
representative model of a edge-on flat pseudo-disk disk with infall and rotation  (see text for further details), is 
shown in red contours, using the same intensity levels as for the data. Grey rectangle shows the velocity 
and angular resolution, and the dashed vertical line denotes the systemic velocity, $6$\,\kms.
}
\label{diskmod}
\end{figure}

At small offset velocities ($\pm10$\,\kms) relative to the systemic velocity, we find a 
structure that is kinematically different from the HVO. We find that the intensity peaks at 
2.7 and 10.9\,\kms~in the $^{12}$CO J=2-1 line profile (Fig.\,\ref{co13co-spec}a), are separated spatially by $0\farcs78$ along an axis at 
$PA\sim$40\arcdeg, in our highest-resolution data (i.e., the ext datacube, with a beam of $1\farcs34\,\times\,0\farcs81$ at $PA=81.46\arcdeg$). A 
position-velocity plot of a cut along this axis shows that the 
low-velocity emission is suggestive of differential rotation (Fig.\,\ref{diskmod}), 
such that the north-east (NE) side is moving towards us, and the SW side is moving away 
from us. This gradient is opposite to that seen for the HVO emission. We label the 
structure producing this emission as the pseudo-disk (We present a model for 
the pseudo-disk in $\S\,\ref{txt:diskmod}$). We infer that the axis of this pseudo-disk, 
orthogonal to the direction of maximum velocity gradient above, is oriented at $PA\sim$130\arcdeg.

However, the kinematics of this 
structure is likely more complex than simply differential rotation, since in the latter 
case, the separation of the red- and blue-shifted emission should become  
progressively smaller for increasingly larger offset velocities from the central velocity, but in 
the PV-plot shown in Fig.\,\ref{diskmod} there appears to be a net separation even at 
the highest velocities. 


A position-velocity plot of a cut along the pseudo-disk major axis extracted from the $^{13}$CO 
J=2-1 ext datacube (not shown) also shows a spatially compact emission structure extending over a 
velocity range $-4 < V_{lsr} < 15$\,\kms~that appears similar 
to the structure seen in the $^{12}$CO J=2-1 PV plot; we conclude that this emission also 
comes from the pseudo-disk.




\subsubsection{Shock and High-Density Tracers}\label{txt:sma-so}
\begin{figure}[htbp]
\includegraphics[angle=0,origin=c,scale=0.95]{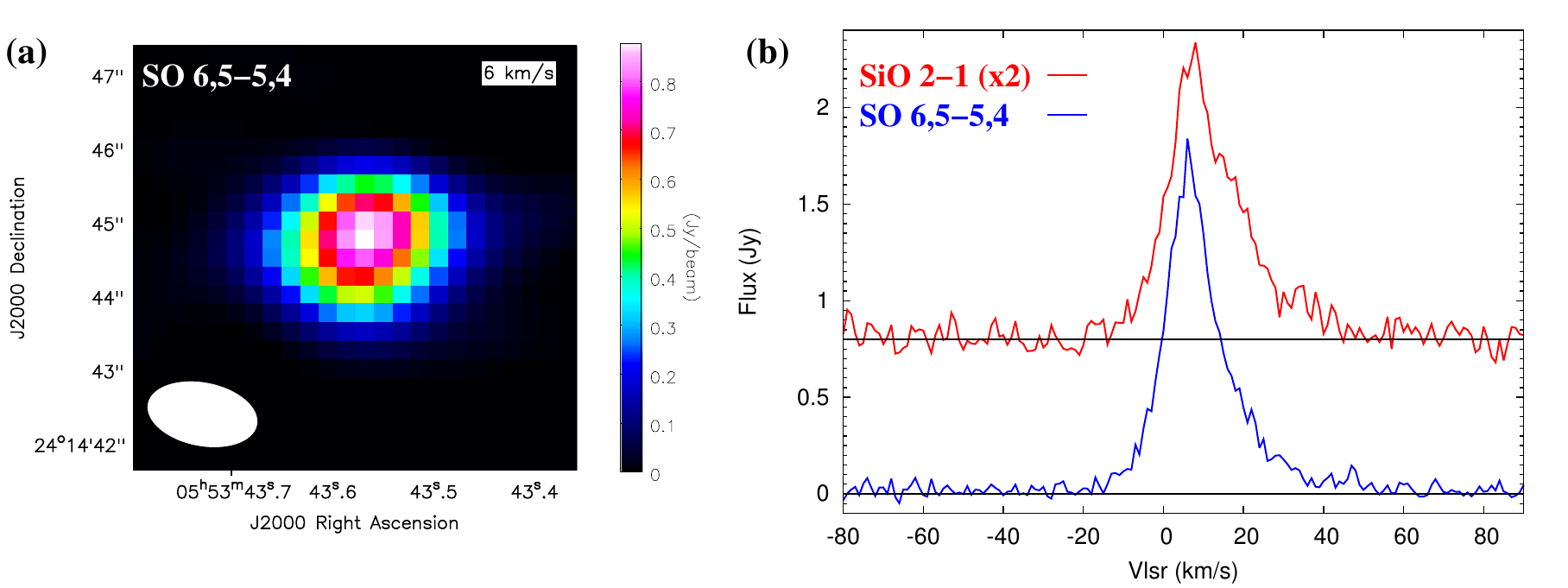}
\vskip 0.5cm
\caption{(a) SO N,J=6,5--5,4 map of IRAS\,05506 obtained with the SMA (ext datacube) at the peak emission velocity, $V_{lsr}=6$\,\kms~(the beam FWHM is 
$1\farcs5\times0\farcs85$, $PA=78.5\arcdeg$ (white ellipse). (b) SO N,J=6,5--5,4 and SiO v=0, J=2-1 spectra extracted from the ext datacube using 
apertures of size $3\farcs6\times3\farcs0$ and $3\farcs9\times2\farcs9$, respectively (i.e., twice the FWHM source-sizes in each line).
}
\label{so-sma}
\end{figure}

The SO N,J=5,6-4,5 map at the peak emission velocity ($V_{lsr}=6$\,\kms)\footnote{a map of the emission integrated over the line-width is similarly compact 
(not shown)}, shows that most of the emission arises from a compact source -- the FWHM size of the core in the SO ext image (Fig.\,\ref{so-sma}a) 
is $\sim1\farcs8\times1\farcs5$, $PA\sim100\arcdeg$). The emission peak is located at coordinates (J2000 RA, Dec): 5:53:43.565, 24:14:44.68. 
Hence, given the FWHM beam-size 
$\sim1\farcs46\times0\farcs85$ ($PA=82.5\arcdeg$), the SO source is marginally resolved, with a deconvolved size of about $\sim1\farcs05\times1\farcs2$. 
The SiO 5-4 (v=0) emission map, shows similar properties as SO N,J=5,6-4,5, but is weaker and therefore has lower signal-to-noise.
The SO and SiO line profiles extracted from apertures twice the FWHM source sizes in these lines  (thus capturing almost all of the source flux), 
have roughly triangular shapes, with clearly visible wings covering about 60\,\kms~(from $V_{lsr}\sim-20$ to 40\,\kms) (Fig.\,\ref{so-sma}b).

\begin{figure}[htbp]
\includegraphics[angle=0,origin=c,scale=0.5]{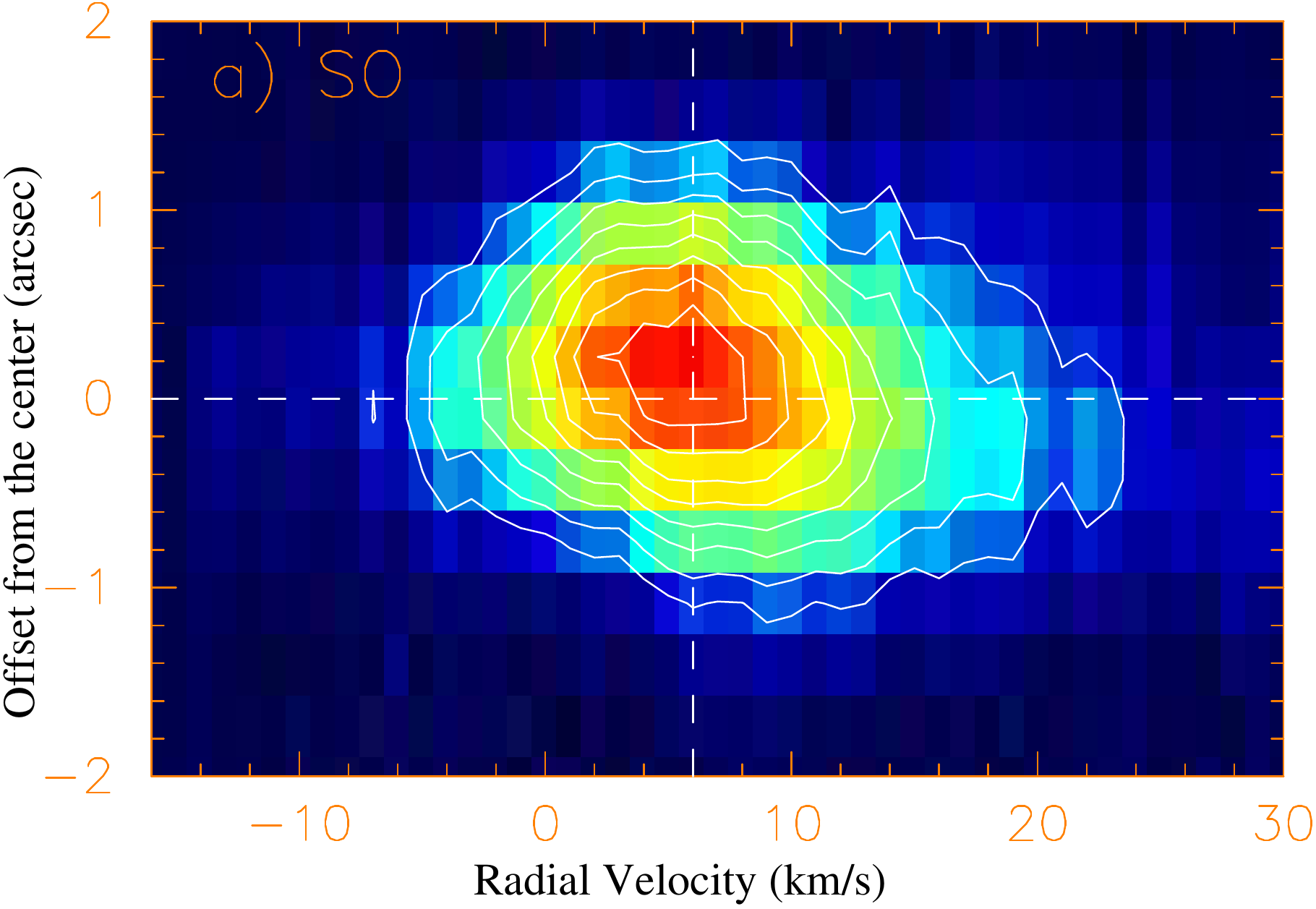}
\vskip 0.5cm
\caption{Position-velocity plot of the intensity along a cut at $PA=40$\arcdeg~in the SO N,J=5,6-4,5 ext datacube. 
Minimum contour level (spacing) is 20 (10)\,\% of the peak intensity. Negative (positive) postion offsets are north-east (south-west) of center.
}
\label{pv_sosio}
\end{figure}

Position-velocity plots of the intensity along a cut at $PA=40$\arcdeg~in the SO N,J=5,6-4,5 ext maps show weak evidence for a velocity 
gradient along $PA\sim40$\arcdeg~over the velocity range $-6<V_{lsr}<20$\,\kms~(Fig\,\ref{pv_sosio}), 
which has the same sense as that seen in the pseudo-disk. We therefore conclude that some fraction of the  
emission seen in the SO line in the velocity range $-6\gtrsim V_{lsr} \gtrsim20$ arises in the pseudo-disk.

The SO and SiO profiles are asymmetric about the central peak with 
the red-shifted side being brighter and broader than the blue one.
Such a shape is a characteristic signature of self-absorption in an optically-thick outflow with a radially-decreasing temperature distribution. We propose 
that some or all of the high-velocity wing emission, i.e., at velocities offset by 
$\gtrsim12-15$ from the systemic velocity arises in an outflow (hereafter ``medium-velocity outflow" or MVO). We do not 
find a convincing velocity gradient for this outflow, since the SO emission is only marginally resolved in our maps. 

The asymmetry of the line profiles from the MVO is opposite to that seen for the emission from the HVO -- for the latter, the \codos~emission from the 
blue-shifted component ($V_{lsr}<-20$\,\kms) is stronger than the red-shifted one ($V_{lsr}>40$\,\kms)
(see Fig.\,\ref{co13co-spec}b). The \codos~profile, in the velocity range of the MVO, has the same asymmetry 
as seen in the SO and SiO lines. This difference in profile asymmetries between the MVO and HVO strengthens our inference that these two outflows 
are distinct kinematic components.


The SO N,J=5,6-4,5 flux measured with the SMA is consistent with that measured with the SMT. We infer that the 
emission seen in the single-dish spectra of the SO arises in the pseudo-disk and MVO.
\subsubsection{Continuum}\label{smaconttxt}
\begin{figure}[htbp]
\hskip -1.0cm
\includegraphics[origin=c,scale=0.95]{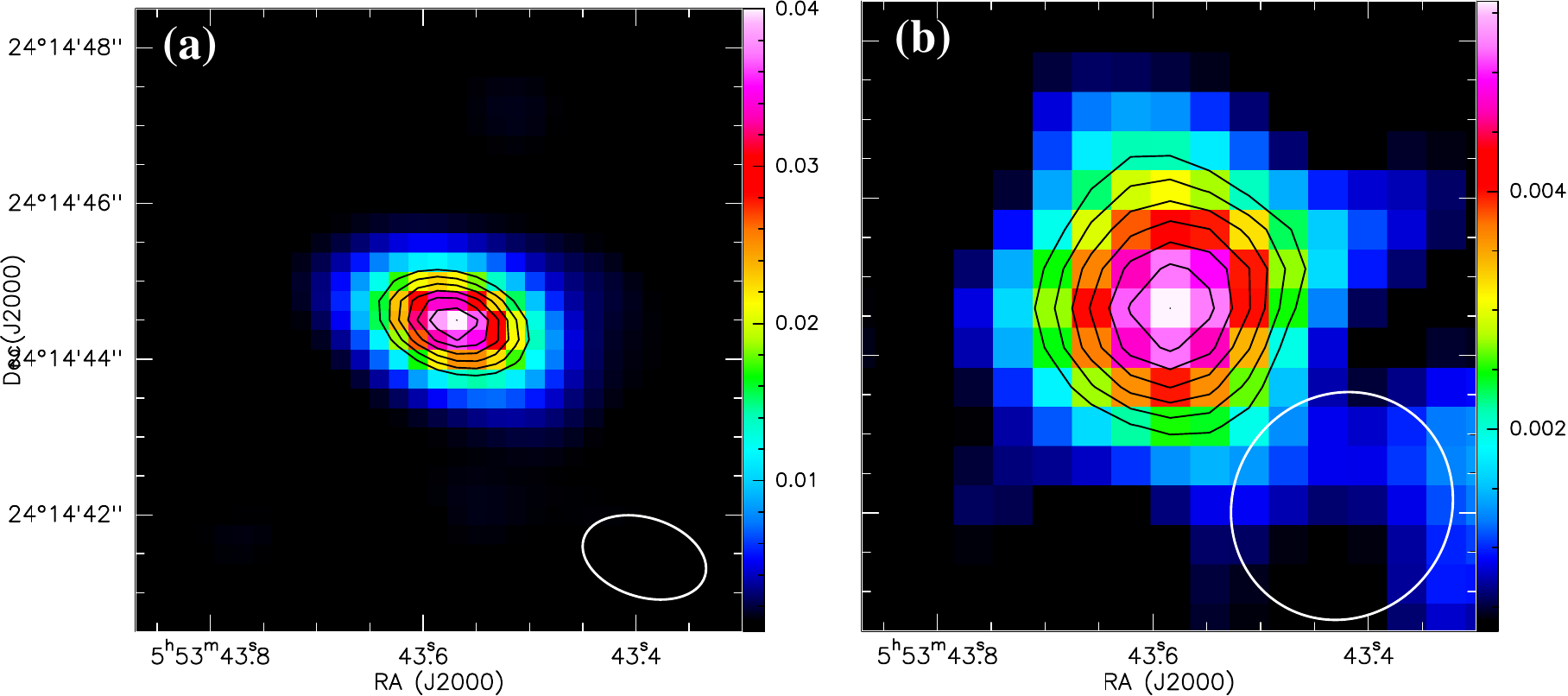}
\vskip 0.5cm
\caption{Millimeter-wave continuum maps of IRAS\,05506 at (a) 1.3\,mm (SMA); beam (white ellipse) is $1.620{''}\times 1.012{''}$ at $PA=72.93\arcdeg$, 
(b) 2.6\,mm (OVRO); beam (white ellipse) is $2.959{''}\times 2.776{''}$ at $PA=-33.37\arcdeg$. Minimum contour level (spacing) is 40\% (10\%) of the peak intensity, 
40.1\,mJy\,beam$^{-1}$ (5.65\,mJy\,beam$^{-1}$) at 1.3 (2.6)\,mm.
}
\label{contmm}
\end{figure}

The 1.3\,mm continuum integrated over line-free channels, shows a largely unresolved 
source, located at  coordinates (J2000 RA, Dec): 5:53:43.570 24:14:44.47 (Fig.\,\ref{contmm}a), with a FWHM size ($1\farcs7\times1\farcs1$), and $PA$ 
($\sim73\arcdeg$) that is very similar to that of the beam ($1\farcs6\times1\farcs0$, 
$PA=73\arcdeg$). Hence, the intrinsic source size is $\lesssim0\farcs47$. The 2.6\,mm continuum source is 
also unresolved (Fig.\,\ref{contmm}b) (beam $3\farcs0\times2\farcs8$).

The total flux 
at 1.3\,mm is $\sim$50\,mJy (45 mJy at a frequency of 220.4\,GHz from the lower sideband, 
and 56 mJy at a frequency of 230.5\,GHz from the upper sideband). Comparing this with the 
continuum flux at 2.6 mm of $5\pm1$mJy measured by Setal08, we 
derive a power-law dust emissivity exponent of 1.4, in the Rayleigh-Jeans approximation. 



\section{The Structure of IRAS\,05506}\label{struct}
The structure of IRAS\,05506 and its environment is complex. However, as a result of being able to map several different molecular lines at different angular 
resolutions, we are able to identify the major components and derive their physical properties.
Many of the derived values are sensitive to the adopted distance, and we have modified 
the 2.8\,kpc distance to IRAS\,05506 adopted in Setal08. We use $V_{lsr}=6$\,\kms~to determine a 
distance of 3.7\,kpc. This value is similar to that derived by Lumsden et al. (2013), 
as part of the Red MSX Source Survey of massive protostars.

Note, however, because of the typical intrinsic velocity dispersion in molecular clouds, the uncertainty 
in the distance to IRAS\,05506 is relatively large. We show below that proper motion measurements 
of IRAS\,05506's high-velocity bullets can be used 
to determine its expansion-parallax distance, with each bullet providing an 
independent measure of the distance. 

\begin{figure}[htbp]
\includegraphics[angle=270,origin=c,scale=0.5]{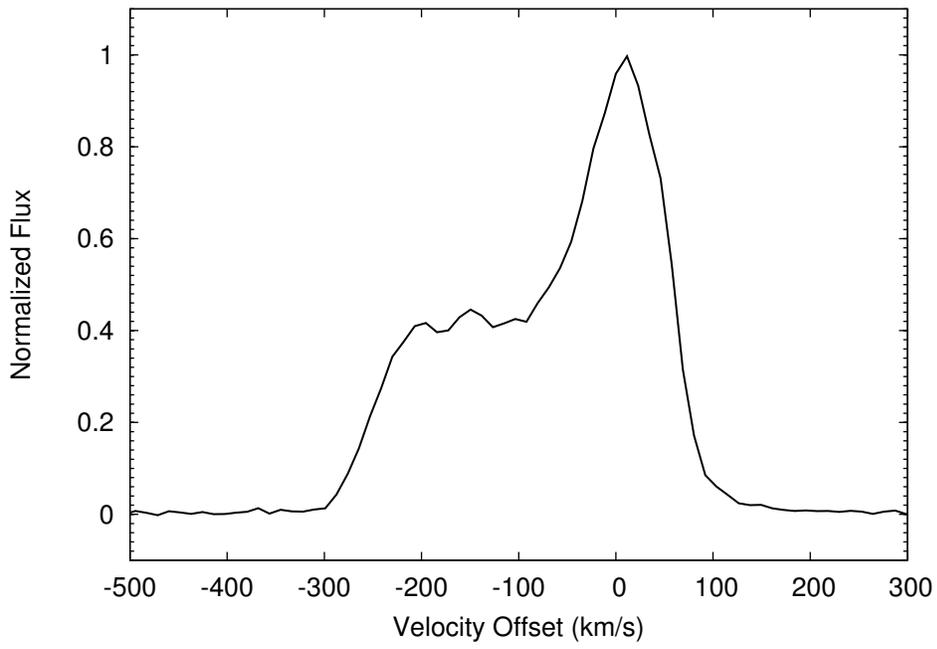}
\caption{The H$\alpha$ velocity profile of knot K1 in IRAS\,05506's bullet spray, from the optical 
spectroscopy in Setal08.}
\label{haknotk1}
\end{figure}

The expected proper motion of the knots is 22.7 mas\,yr$^{-1}$\,sin\,$i$\,($V_k$/400\,\kms)\,(3.7\,kpc/D), 
where $V_k$ is the 3D velocity of the knot, and $i$ is 
the inclination of the axis, relative to the 
line-of-sight, along which each knot is moving. Both $V_k$ and $i$ can be estimated from the 
radial velocity profiles of the bow-shock emission associated with each knot. Bow-shock models show that for $i$=90\arcdeg, the velocity structure 
is symmetric around the systemic velocity, and becomes increasingly asymmetric as $i$ becomes smaller or larger than 90\arcdeg~(see Fig.\,1 of Hartigan et al. 
1987: HRH87); the velocity width of the profiles at the base is equal to the 3D velocity of the shocked material at the 
apex of the bow-shock, i.e., $V_k$. 
We have extracted the H$\alpha$ velocity profile of knot K1 in IRAS\,05506's bullet spray, from the optical 
spectrsocopy in Setal08. The observed line profile, with a narrow peak near the systemic velocity, and a broad shoulder extending towards 
blue-shifted velocities (Fig.\,\ref{haknotk1}), appears to closely resemble the $i$=60\arcdeg~model profile for a shock velocity $V_s=400$\,\kms, including 
its width at the 
base, $\sim400$\,\kms (panel N in Fig.\,1 of HRH87)\footnote{although the profile shape in panel R also looks similar, its width at the base 
($\lesssim300$\,\kms) is 
significantly smaller than observed}. We conclude that $i\sim$60\arcdeg~for knot K1, and that $V_k=V_s=$400\,\kms. Thus 
the total proper-motion from our 1st-epoch imaging in 2002 Oct to (say) a 2nd-epoch 
in 2018, is 363\,mas\,(3.7\,kpc/D), and thus easily measurable, even for D as large as $\sim$15\,kpc. 
Conservatively accounting for uncertainties in $i$ and the proper motion, the expansion-parallax distance to IRAS\,05506 can 
be determined with an uncertainty of about $\pm$10\%.

\begin{figure}[htbp]
\includegraphics[origin=c,scale=0.9]{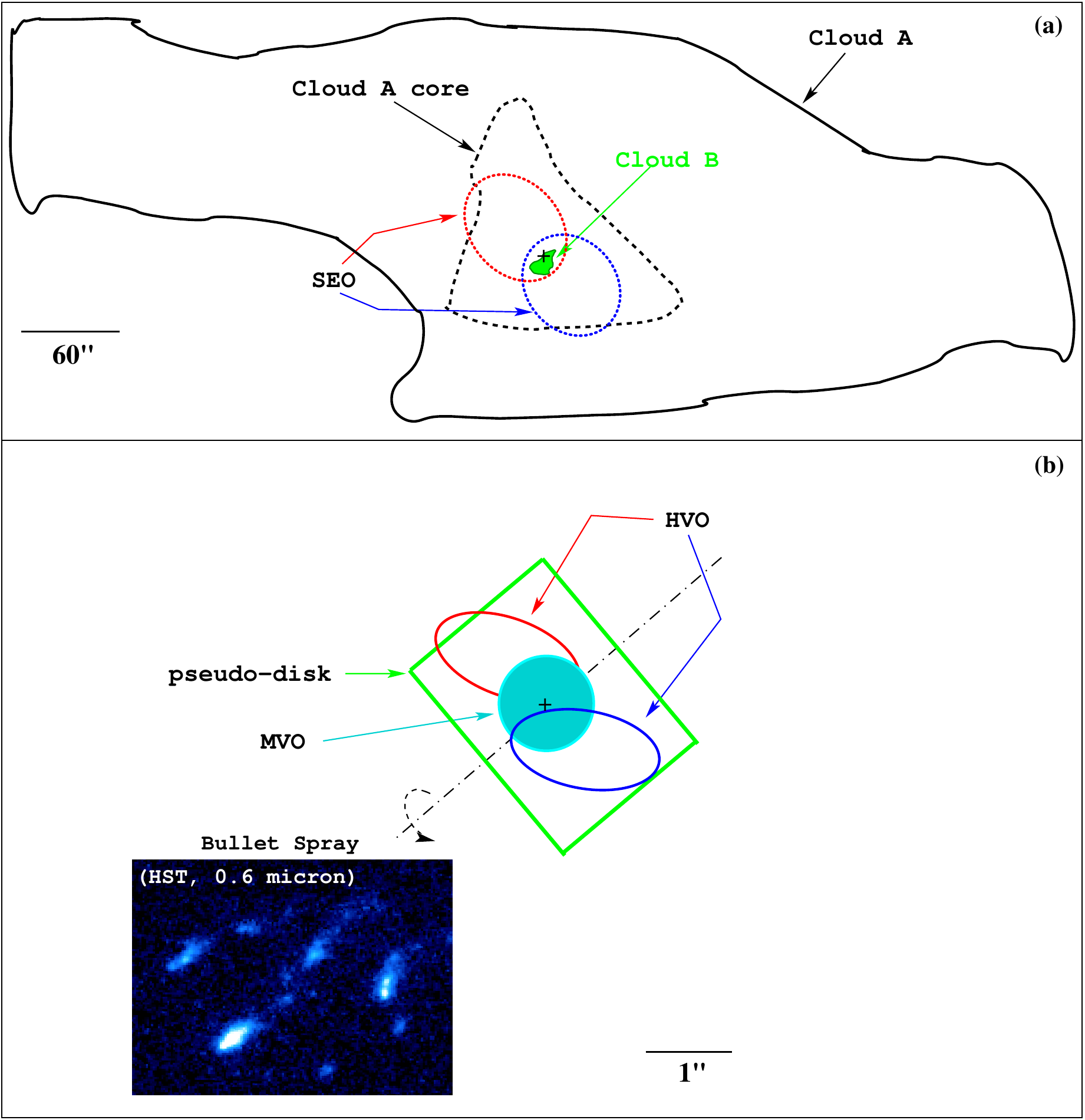}
\caption{Schematic showing the various structural components of IRAS\,05506 and its extended environment. The sizes of the various extended cloud components,  
the pseudo-disk, and the separations of red- and blue- shifted lobes of the HVO and SEO are roughly to scale. The sizes and shapes of the red- and 
blue- shifted lobes of the SEO and HVO are only indicative. 
(a) The very extended cloud A, its dense core (cloud A-core), cloud B, and the slow extended outflow (SEO); the location of IRAS\,05506 Sa is marked with 
a cross.  (b) The 
high-velocity outflow (HVO), the medium-velocity outflow (MVO), and the edge-on pseudo-disk (all of which lie inside cloud B), 
associated with Sa (black cross). 
Inset in panel $b$ shows the 
high-velocity bullet spray found by Setal08 at their actual location relative to the central source. The rotation axis and sense of rotation of the pseudo-disk 
are shown by the black dash-dot line and arrowed arc. 
}
\label{schematic}
\end{figure}

The  major components of IRAS\,05506 and its environment are shown schematically in Fig.\,\ref{schematic}, and described below. 
Some salient properties of these components are tabulated in Tables\,\ref{tab:phys_parm} and \ref{tab:phys_prop} for easy reference.

\subsection{Large-Scale Structure}
\subsubsection{Cloud A and its Dense Core}
Our single-dish $^{12}$CO J=2-1 (e.g., Fig.\,\ref{co21map-smt}) and J=1-0 (Setal08) maps reveal that 
IRAS\,05506 is located towards a very extended cloud (cloud A) with 
$V_{lsr}=6$\,\kms. The ``absorption" feature that is seen in the $^{12}$CO J=2-1 and 1-0 
single-dish spectra of IRAS\,05506 at $V_{lsr}=6$\,\kms~is presumably due to relatively 
cold gas in this cloud that is optically thick these lines. The brightness temperatures of the line emission in these spectra at 
$V_{lsr}=6$\,\kms~(4\,K and 2.5\,K, respectively, after convolving the $^{12}$CO J=2-1 data to the 
larger beam-size of the $^{12}$CO J=1-0 data) indicates an average kinetic temperature 7\,K in Cloud A. 

As noted previously (\S\,\ref{smtco},\ref{spire}), the $^{13}$CO J=2-1 map delineates a dense core (see Figs.\,\ref{13co21map-smt}, \ref{i05506herschel}) 
within cloud A. 
We make the reasonable assumption that the single-dish profiles of $^{13}$CO (J=1-0 and 2-1) and C$^{18}$O (J=1-0), all of which are 
centrally-peaked at $V_{lsr}=6$\,\kms, arise from this dense core. The ratio of the single-dish $^{13}$CO J=1-0 to 
C$^{18}$O J=1-0 spectra is 6.3, close to the local interstellar $^{13}$CO to C$^{18}$O abundance ratio of, [$^{13}$CO]/[C$^{18}$O]=6.1 (Wilson 1999), 
from which we infer that the $^{13}$CO J=1-0 line 
in the cloud A core, averaged over a 
$1\arcmin$ scale, has an optical depth of slightly less than unity ($\sim0.7$), and the 
kinetic temperature is about 9.5\,K. The beam-averaged column density of C$^{18}$O is 
N(C$^{18}$O)=$6\times10^{14}$ cm$^{-2}$, implying a molecular hydrogen column density, 
N(H$_2$)=$3.5\times10^{21}$ cm$^{-2}$, assuming a standard ISM C$^{18}$O/H$_2$ abundance 
ratio of $1.7\times10^{-7}$ (Goldsmith, Bergin, \& Lis 1997).

\subsubsection{Cloud B}
Our SMA $^{13}$CO J=2-1 map resolves out the smooth extended emission from cloud A and its core, and 
shows a roughly trapezoidal-shaped cloud (cloud B), extending about $14{''}$ (at its 
half-intensity countour) along $PA\sim135\arcdeg$ (Fig.\,\ref{13co21cloud}a), i.e., 
roughly along the axis of the wide-angle bullet spray. The C$^{18}$O J=2-1 map shows a 
roughly similar shape (Fig.\,\ref{13co21cloud}b). A moment 0 map of the $^{13}$CO J=2-1 
emission over the velocity range V$_{lsr}=-4.5$ to 18.5\,\kms~shows the outflow source in 
IRAS\,05506 located near the NE periphery of this cloud. Emission from cloud B is seen in 
the $V_{lsr}=5-7$\,\kms~range, peaking between channels centered at $V_{lsr}=6$ and 
$V_{lsr}=7$\,\kms.

Assuming that the 
densities are high enough for 
the lines to be thermalized, and [$^{13}$CO]/[C$^{18}$O]=6.1 (as above), we find 
that the $^{13}$CO J=2-1 emission is optically thick 
everywhere in cloud B, by fitting the brightness temperatures of the $^{13}$CO to 
C$^{18}$O J=2-1 emission in the SMA maps. Towards the location of 
IRAS\,05506, the $^{13}$CO J=2-1 optical depth is, $\tau_{13CO21}=4$ and the kinetic temperature is, $T_{kin}=8.5\,K$. In comparison, 
$\tau_{13CO21}=7$ and $T_{kin}=11\,K$ in Pk$_{b}$, located 5\farcs3 south of IRAS\,05506.

\subsubsection{Cloudlet B1}
The nature and association of Cloudlet B1 with the IRAS\,05506 outflow source is not clear. For now, all we can say is that it 
produces additional features in some of the maps and spectra, adding to their complexity. For example, it is likely responsible for the emission peak near 
$V_{lsr}=7.5$\,\kms~seen in 
the single-dish profiles of $^{12}$CO J=3-2 and 4-3, and the prominent peak at 
$V_{lsr}=7-8$\,\kms~in the $^{12}$CO J=2-1 SMA extsub line profiles. This component was 
shown to be modestly extended relative to the SMA extsub beam, but since the ratio of 
the $^{12}$CO J=4-3 to the J=3-2 peak intensity is about 1.8, it must be significantly smaller  
than the $^{12}$CO J=4-3 beam size ($15{''}$), in order for beam-dilution to account for 
this ratio. 

\subsubsection{SEO}\label{seoprops} 
The SEO, extends  over about 1 pc, and is seen in our single-dish \codos~J=2-1 maps.
The age of the SEO, estimated by dividing the separation between 
its blue- and red-shifted lobes, of $\delta_{SEO} = 51\farcs8$ (as seen in the $V_{lsr}$ velocity ranges 
$(0,4)$\,\kms~and $(8,12)$\,\kms~(Fig.\,\ref{co21-smt-blue-red})), by the average 
velocity separation of 8\,\kms~between these lobes, is $t_{SEO}\sim114,000$\,yr.

We estimate a total \codos~J=2-1 flux of 22.9 and 28.8 Jy--\kms~in the blue and red lobes of the SEO, which, assuming a nominal kinetic temperature 
of 15\,K and optically-thin emission, implies a total mass, scalar momentum 
and kinetic energy of 1.6\,\ms, 4.7\,\momunit, and $0.3\times10^{45}$\,\keunit. We have assumed a 
fractional CO abundance $f_{CO}$=[CO]/[H$_2$]=$10^{-4}$ (here and elsewhere in 
the paper). 
Its presence implies that a very significant amount of kinetic energy 
has been added to a $\sim1$\,pc region around IRAS\,05506. 

The sense of the velocity gradient, the axis orientation, as well as the scalar momentum, of the HVO and SEO are similar.  
Hence, it is likely that the SEO has resulted  
from the interaction of the HVO with the surrounding ambient cloud 
material. Thus the value of $t_{SEO}$ derived above 
is an upper limit. A better approximation to the SEO age can be derived assuming a uniform deceleration, in which case, 
$t_{SEO}=0.5\,\delta_{SEO}/(V_{i}+V_{f})=9300$\,yr, where $V_{i}$ and $V_{f}$ are the initial and final velocities of the material in 
the SEO, taken to 
be the average outflow velocity of the HVO (94\,\kms; see \S\,\ref{hvomvoseo} below) and the SEO (4\,\kms).


\subsection{Small-Scale Structure}
The central compact source in IRAS\,05506 as seen in our high-angular rsolution molecular-line maps, has several major structural components that 
can be identified by their spectral signatures (Fig.\,\ref{co13co-spec}) and spatial 
extents -- we describe these below.

\subsubsection{The Pseudo-Disk}\label{txt:diskmod}
The pseudo-disk is seen in $^{12}$CO J=2-1 emission at $2.7 < 
V_{lsr} < 10.9$\,\kms, with its axis aligned roughly along that of the wide-angle bullet spray. 
We estimate a rough upper limit to the optical depth of the 
pseudo-disk emission at the velocity of the blue peak by comparing the average intensities of 
the \codos~J=1-0 ($\sim10$\,K) and J=2-1 emission ($\sim20$\,K) in spectra extracted by 
averaging over a $3{''}\times3{''}$ region, after convolving the \codos~J=2-1 map to the 
same beam size as the J=1-0 one. We find that the $^{12}$CO J=2-1 optical depth is $\sim 
1$, implying an optical depth for $^{13}$CO J=2-1 that is a factor 60 lower. 

We have made a simple squat-cylindrical, edge-on model of the pseudodisk, assuming that it 
has both a free-fall velocity (i.e., radial) component and rotation, with a conservation 
of angular momentum, using a code described in Lee et al. (2014). We have therefore assumed that 
$V_{\phi}(r)=V_{0,\phi}\,(r/1{''})^{-1}$, and $V_r(r)=-V_{0,r}\,(r/1{''})^{-0.5}$, and 
adjusted $V_{0,\phi}$ and $V_{0,r}$, in order to obtain a good fit to the contours in the PV 
plot shown in Fig.\,\ref{diskmod}, as well as the observed brightness. The temperature 
is assumed to be $T_{exc}=T_1\,(r/1{''})^{-0.4}$, and the molecular hydrogen density, 
$\rho=\rho_1\,(r/1{''})^{-0.5}$. The temperature power-law index, $-0.4$ is assumed to be the same 
as in low-mass envelopes (e.g., Hogerheijde 2001, Lee et al. 2006). 
The density power-law exponent, $-0.5$, results in a constant
mass infall rate, given our assumption of a fixed disk thickness and $-0.5$ for the 
power-law exponent of the infall velocity above. The parameters $T_1$ 
and $\rho_1$ are adjusted to reproduce the absolute 
intensity of the \codos~J=2-1 emission from the pseudo-disk. All models have a disk 
thickness of $2{''}$, and an outer disk radius of $1\farcs3-1\farcs5$ (the inner 
radius is assumed to be much smaller than the outer radius.) The model PV plot is 
generated after convolving the model brightness distribution with the beam.

We find good fits with $V_{0,r}=2-3$\,\kms, $V_{0,\phi}=4$\,\kms, $\rho_1=2.5\times10^5$ 
cm$^{-3}$, and $T_1=65-75$\,K. A representative model is shown in Fig.\,\ref{diskmod}. The 
central stellar mass is, $M_{*}=(8.3-18.8) (D/3.7\,kpc)$\,\ms, and the disk mass is $M(disk,tot)= 
(0.64-0.85) (D/3.7\,kpc)^2$\,\ms. This mass range is consistent with the mass of the young star associated with IRAS\,05506, Sa, 
derived from SED modelling (Setal08: 11.2\,\ms~at D=3\,kpc), and 
scaled to our distance of 3.7\,kpc. 

The compact continuum source reported above (S\,\ref{smaconttxt}), is likely to be thermal dust emission from the dense, inner 
region of the pseudo-disk.   
Without a detailed knowledge of the dust density structure, it is not possible to build a robust model of 
the thermal dust emission, although we can derive some rough constraints from the observed size of the continuum source. 
The characteristic dust emission radius, assuming no extinction or reddening of the starlight, is 
given by $r_d=[\frac{L_{*}T_{*}^\beta}{16\pi\sigma}]^{1/2}T_d^{-(2+\beta/2)}$, where $\beta$ is the 
dust emissivity exponent (Sahai et al. 1999). Since $r_d\lesssim0\farcs47$, 
then $T_d\gtrsim185$\,K. We have taken $T_{*}=4700$\,K and a luminosity scaled up to D=3.7\,kpc, $L_{*}=8900$\,\ls, from Setal08. 

However it is likely that there is significant absorption of the starlight by intervening 
dust close to the star. If we assume that, say, 20--50\% of the starlight is re-radiated at an 
effective temperature of $T_{*}=650-1000$\,K by a compact source of optically-thick dust surrounding the central star, 
then $T(disk)_{dust}\gtrsim(84-111)$\,K in order that $r_d\lesssim0\farcs47$. 

From our measured 1.3\,mm dust continuum flux, the above dust temperature estimate, and a dust emissivity (per unit dust mass) of 
$\chi_{\nu}=$1.5\,cm$^2$\,g$^{-1}$ at 1.3\,mm, we derive 
a dust mass and total mass of $M(disk,dust)\lesssim(0.013-0.017)$\,\ms~and 
$M(disk,tot)\lesssim(1.3-1.7)$\,\ms~(assuming the standard gas-to-dust ratio of 100). This disk mass is in 
reasonable agreement with that derived from our pseudo-disk kinematical model, given the uncertainties in the 
absolute value of the dust emissivity and our very simple thermal dust model.

\subsubsection{Outflows}\label{hvomvoseo}
There are three outflows emanating from the lcation (or near vicinity of) IRAS\,05506 Sa -- the SEO, the HVO, and the MVO. The physical 
properties of the SEO  have been 
derived earlier (\S\,\ref{seoprops}); those for the HVO and MVO properties are derived below.

\paragraph{HVO} The HVO is seen most clearly in $^{12}$CO J=2-1 emission at 
$V_{lsr}\lesssim -20$\,\kms and $V_{lsr} \gtrsim 40$\,\kms, and is aligned along $PA\sim40$\arcdeg, with the 
blue- (red-) shifted component SW (NE) of the center. The HVO axis is thus roughly orthogonal to that of the wide-angle spray (121\arcdeg--155\arcdeg: 
Setal08). This 
outflow was already noted by Setal08 in their $^{12}$CO J=1-0 data, at a somewhat larger but uncertain $PA$ since the spatial 
separation of the blue- and red-shifted components was small compared to the angular resolution of the data 
presented here ($3\farcs9 \times 2\farcs9$).

For this outflow, the \codos~J=1-0 and J=2-1 brightness temperatures in the HVO averaged over a 5.5\arcsec~box, 
are rather low -- e.g., about 0.4\,K and 1\,K, respectively in the range $-50<V_{lsr}(\kms)<-20$, and about 0.1\,K and 0.59\,K in the 
range $40<V_{lsr}(\kms)<70$. For larger velocity offsets from the systemic velocity, the brightness temperatures are even lower.
The ratio of the J=1-0 to 2-1 brightness temperature, $R_{12}$, is about 0.34 in the 
40 to 110\,\kms range (i.e., for the red-shifted outflow), and 
0.47 in the -20 to -100\,\kms range (i.e., for the blue-shifted outflow), demonstrating that emission from both lines is optically thin. 

The age of the HVO, estimated by dividing the separation of $1\farcs4$ between 
blue- and red-shifted lobes of the HVO as seen in the velocity ranges $(-116,-84)$\,\kms~and $(72,104)$\,\kms~(Fig.\,\ref{extremevel}), by the average 
velocity separation of 188\,\kms~between these lobes, is $t_{HVO}\sim130$\,yr.

The total \codos~J=1-0 and 2-1 flux from the high-velocity blue-shifted (red-shifted) outflow component is 
9.8 (2.5) and 80 (30)\,Jy-\kms, from which we estimate that $T_{kin}\sim20 (30)$\,K, and the  
mass, scalar momentum and kinetic energy are about $0.052$ ($0.022$)\,\ms, 7.2 (2.6)\,\momunit, and $6.1\times10^{45}$ ($1.7\times10^{45}$)\,\keunit, giving 
a total mass, scalar momentum and kinetic energy for the HVO of 0.074\,\ms, 9.8\,\momunit, and $7.8\times10^{45}$\,\keunit.

\paragraph{MVO}\label{txt:mvo} 
The MVO is seen most prominently in emission in the extreme wings of the SO\,N,J=5,6-4,5 and
SiO\,J=5-4 lines in the SMA data. This outflow is relatively compact, hence we have used the ext datacube for our 
subsequent analysis of this line.

We estimate the age of the MVO as follows. We average the moment 0 maps of the blue-wing (i.e., $-20<V_{lsr}(\kms)<-6$) and red-wing 
(i.e., $20<V_{lsr}(\kms)<40$) emission, and determine its size to be  
$1\farcs6\times1\farcs3$  (FWHM). Deconvolving this with the beam, $1\farcs46\times0\farcs85$, we obtain a deconvolved size of $0\farcs8\times1\farcs0$. We 
divide the 
mean size by the average velocity separation between the blue and red-velocity intervals, i.e., 43\,\kms, giving an age 
$t_{MVO}\sim360$\,yr. The (relatively uncertain) age of the MVO is thus comparable to that derived for the knots in the wide-angle spray by Setal08: e.g., 265\,(D/3.7 kpc) 
and 225\,(D/3.7 kpc) for knots K1 and K2, suggesting that these outflows are associated with the same triggering event. 

Convolving the \codos~J=2-1~extsub datacube to the same angular resolution as the \codos~J=1-0 datacube, we find that the flux ratio of the \codos~J=1-0 to 2-1 
in the MVO, is about 0.1 for the blue-shifted component, and 0.075  
for the red-shifted component. Modeling these ratios shows that emission from both lines is optically thin, and that the kinetic temperature lies 
in the $20-30$\,K range. From the total \codos~J=2-1 flux of 8.3 (14.7)\,Jy-\kms~for 
the blue-shifted (red-shifted) component of the MVO extracted from an aperture equal to the FWHM size of the MVO, 
we estimate the mass, scalar momentum and kinetic energy\footnote{since the fractional flux within an aperture of size equal to the FWHM of a Gaussian 
distribution is 0.5 of the total} to be 
about 0.011 (0.02\,\ms, 0.21 (0.44)\,\momunit, $0.78\times10^{44}$ ($2.1\times10^{44}$)\,\keunit. 
The total mass, scalar momentum and kinetic energy in the MVO is thus 0.031\,\ms, 0.65\,\momunit, $2.9\times10^{44}$\,\keunit.



We can also use the SO data to further constrain the properties of the shocked gas in the MVO. The peak 
intensity of the SO N,J=5,6-4,5 source in the SMA extended array data is 1\,Jy\,beam$^{-1}$, implying a brightness temperature of 26\,K, 
and since the source is marginally resolved at best, this should be considered a lower limit. 

Since the broad emission in the four SO lines observed with the ARO (Fig.\,\ref{fig:smt-12m-so-hco}) likely arises in the MVO (S\,\ref{txt:sma-so}), we can use 
the relative ratios of the integrated 
fluxes in the velocity ranges for the MVO component for these lines (corrected for beam-dilution) to constrain the 
density and temperature of shocked gas in the MVO. As these data have modest S/N, we 
only use the red component of the MVO, which appears to be stronger than the blue one, to compute these ratios. We find that 
the uncorrected ratios for the four SO lines 1, 2, 3, and 4 (\S\,\ref{txt:smtso}), are 0.26:0.26:0.51:1. 
The ratios, corrected for beam-dilution, are 0.69:0.51:0.72:1, assuming the source-size 
is significantly smaller than the single-dish beams. We have used the non-LTE RADEX code  (van der Tak et al. 2007) 
to compute the line ratios for a variety of kinetic temperatures and densities, and we 
find that only for densities $\rho(H_{2})\gtrsim10^7$\,cm$^{-3}$, does the 
excitation temperature of the 5,6-4,5 and 6,7-5,6 lines approach the kinetic temperature; at this density, we find that $T_{kin}\gtrsim80$\,K 
for a reasonable fit to the line ratios (for lower values of $T_{kin}$, the model intensities of the lower-frequency pair, N,J=3,4-2,3 and 
N,J=4,3-3,2, become too large relative to the higher frequency pair, N,J=6,7-5,6 and 
N,J=5,6-4,5). Fits can be obtained at somewhat lower densities as well, by increasing the kinetic 
temperature -- the lowest viable value of the density is $\rho(H_{2})\sim10^6$\,cm$^{-3}$, for which the temperature is $T_{kin}\sim120$\,K. 
The significantly higher temperatures derived from the SO lines compared to the CO lines implies that the former is tracing the hot, shocked fraction of 
the material in the MVO.

\section{Discussion}\label{discus}
The closest counterpart to the wide-angle bullet spray that Setal08 found towards IRAS\,05506 is the 
explosive BN-KL wide-angle outflow in OMC1. Furthermore, in IRAS\,05506, the HVO is directed along an axis 
orthogonal to that of the bullet-spray, a likely counterpart to the collimated outflow that  
emerges from Source I in BN/KL (Plambeck et al. 2009, Zapata et al. 2012) along an axis that is orthogonal to the wide-angle outflow as 
seen in $H_2$ and [FeII] line emission (Allen \& Burton 1993, BCM11). However, in addition to these similarities, 
we note that IRAS\,05506 shows three additional components, 
namely the pseudo-disk, the MVO and the SEO, for which there are no clear counterparts in BN/KL. 

The wide-angle bullet spray shares a common axis with the pseudo-disk, and has an age comparable to 
that of MVO, suggesting that these three structures are intimately linked together. The rotation of the pseudo-disk may be a 
result of the overall rotation of cloud B for the following reasons. 
The observed alignment of cloud B's long axis with that of the 
wide-angle outflow is unlikely to be due to the impact of the outflow on Cloud B as there is no 
obvious velocity gradient seen along this axis in cloud B. It is plausible that cloud B's long axis 
is aligned with its rotational axis and 
thus cloud B is the pre-existing dense cloud core in which IRAS\,05506's pseudo-disk was formed, naturally accounting 
for these two structures having a common axis

We note that Cloud B shows several distinct compact sources of dense gas (e.g., $Pk_{a}$, $Pk_{b}$) in emission at 
the systemic velocity $V_{lsr}\sim6$\,\kms~which do not appear to be associated with stars or outflows. In contrast, 
IRAS05506, which is associated with a massive star and multiple outflows, cannot be distinguished from Cloud B emission at the systemic velocity 
$V_{lsr}\sim6$\,\kms. 
Thus, much of the low-velocity gas in the pre-existing density peak where IRAS05506's massive star was born, 
has largely been removed, presumably by the outflows that came into existence after its formation.

We briefly discuss below past and current models for the BN/KL outflows, in order to investigate their applicability 
to IRAS\,05506. A model involving dynamical decay of a 
non-hierarchical system of massive stars was proposed for the BN/KL wide-angle outflow (Bally \& Zinnecker 2005, 
Rodr{\'{\i}}guez et al. 2005, BCM11), with the  
two most massive components of the stellar system being BN (10-13\,\ms~runaway B star) and 
radio Source I (a hypothetical massive binary).
BCM11 proposed that these stars were ejected from their natal cloud core due to a dynamical 
interaction between them, and that the gravitational energy released by the formation of 
binary Source I powers the bullet spray.


An alternative hypothesis was proposed by Tan 
(2004) and Chatterjee \& Tan (2012) in which the BN-KL outflow source is produced by 
the chance encounter of a runaway star (BN) with the KL star-forming core. According to this 
hypothesis, BN was supposedly ejected from the $\theta^1$Ori\,C system 4500\,yr ago and 
its interaction with the KL cloud within the last 1000 years has produced recently enhanced accretion and outflow activity 
(i.e., the bullet spray) from Source I.

The major weaknesses in each of the above models have been summarized in a recent study by Luhman et al. (2017).  
In the dynamical decay model, the mass of $\sim20$\,\ms~for Source I implied by conservation 
of momentum with BN alone is much larger than the dynamical mass of Source I (7\,\ms: Plambeck \& Wright 2016). In the chance-encounter model, 
there are two problems -- first, BN appears to be younger than the Trapezium stars, and second, this model 
does not explain the large motion of Source I. 

Goddi et al. (2011: Getal11) discuss the above models and two additional ones (nos. 1--4 in \S\,6 of their paper). The latter involve a 
close passage between two (initially unbound) protostars, BN and Source I. In the first of these (no. 2), this passage produces enhanced 
accretion in the disk of Source I, but Goddi et al. reject this model, since a binary encounter cannot produce 
stellar ejection, as appears to be the case for BN and Source I. In 
the second one (no. 4), favored by Getal11, Source I was originally a soft, massive (20\,\ms) binary which got hardened as a result of the 
encounter with BN, and the resulting increase in gravitational potential energy powered both the wide-angle spray and 
the ejection of BN and Source I.

Luhman et al. (2017) provide new evidence that resolves the mass discrepancy for Source I mentioned above, 
thus strengthening the dynamical decay model as the 
likely one for the BN/KL explosive outlow. Specifically, they 
find that source x (from Lonsdale et al. 1982) in the KL nebula, which has a high proper motion, was 
involved in the dynamical decay that separated Source I from BN, 
and source x's momentum 
reduces the momentum requirement for Source I, bringing the latter's mass inferred from momentum balance  
into consistency with its dynamical mass. However, the 
relatively large value of the disk mass in Source I ($\sim0.1$\,\ms) is difficult to obtain in 
the dynamical decay model, either from retention of a pre-existing disk or re-formation in the short time that 
has elapsed after the decay ($\sim$500\,yr) (Plambeck \& Wright 2016).

Since we have not identified 
the multiple individual members of a dynamically-decaying stellar cluster that may be associated with IRAS\,05506, 
we cannot yet test the dynamical decay model for this source. We 
note, though, that near-IR HST imaging shows the presence 
of several faint stars (stars fs1, fs2, fs3, fs4 in Fig.\,2 of Setal08) and bright 
stars (e.g., Sb: see Figs.\,2 and 5 of Setal08) within a $\sim5{''}$ radius around the central optical/infrared source of IRAS\,05506, Sa,  
some of which may be members of such a cluster. 
Second-epoch HST imaging observations 
can be used to look for proper motions of  
the stars found in the near vicinity of Sa. If any of the stars around the IRAS\,05506 outflow source was ejected 
from a common center somewhere in the vicinity of IRAS\,05506 Sa as a result of the dynamical decay, 
then from their typical angular offsets of about $2-5{''}$ from Sa, we estimate that the   
corresponding stellar proper motions would be in the range 8--20\,mas\,yr$^{-1}$. The resulting total proper motion of 
$\sim$128-320\,mas over 16\,yr between epoch 1 and a future observing epoch 2, is easily measureable.

The rather long age of the SEO may argue against the dynamical decay model 
of BCM11 for IRAS\,05506. In this model, the collimated outflow from Source I results from a disk that is formed after the 
explosive disintegration event producing the wide angle spray\footnote{The apparent major axis of the disk in Source I (and in BN) is aligned with Source I's 
(BN's) proper motion, and BCM11 explain this as the result of these disks being formed by Bondi-Hoyle accretion onto these stars from the cloud through which 
they are moving, assuming the latter has a significant density or velocity gradient orthogonal to the stellar motion (see last para in Section 4.2 of their 
paper)}. Thus the collimated outflow in Source I is younger than the wide angle spray, i.e., 
$\lesssim500$\,yr. However  
the SEO's age of $\sim9300$\,yr suggests that the HVO (the presumed counterpart to the collimated outflow from Source I), which has 
likely produced the SEO (\S\,\ref{hvomvoseo}), has been in operation for an equally long time. Thus, contrary to the expectation of the 
dynamical decay model, the HVO, and thus the disk that drives it, is much older than the bullet-spray in IRAS\,05506\footnote{An obvious caveat to this line 
of reasoning is that the disk in the IRAS\,05506 source may be long-lived, having survived the violent distintegration process.}.  
If the chance-encounter model does apply to IRAS\,05506, then the origin of the MVO and bullet spray is due to a sudden increase in the accretion rate of the 
pseudo-disk resulting from a chance-encounter of a runaway star with the latter.

A model involving close passage between two protostars (no. 2 of Getal11) may also be applicable IRAS\,05506,
since (unlike BN/KL) we do not have any evidence for a stellar ejection in its case. If so, 
then one of the two protostars in IRAS\,05506 is the massive star associated 
with the pseudo-disk (i.e., Sa), and the other is a star with a disk that produces the HVO. The observed specific orientations of 
the axes of the pseudo-disk and the HVO (which are roughly orthogonal) would then be purely coincidental.

\section{Conclusions}\label{conclude}
We have carried out a study of molecular-line and continuum emission from the interstellar bullet source IRAS\,05506 and its environment using both single-dish 
and interferometric observations. Our major conclusions are listed below:\\
\begin{enumerate}
\item The central optical/IR source Sa in IRAS\,05506 is located at the north-east periphery of a dense, trapezoidal shaped cloud (cloud B) with an extent of 
about  
$14{''}$ (at its half-intensity countour). Cloud B lies within the dense core (Cloud A core, size $\sim2\farcm1$) of a  
large cloud (cloud A, size $\>11\farcm2\times5\farcm2$). The radial velocity of clouds A and B is $V_{lsr}\sim$6\,\kms. 
The Cloud A dense core is prominently 
seen in thermal dust emission at wavelengths in the $250-500$\,\micron~range, as observed with the 
Herschel/SPIRE camera. 
\item A very extended bipolar outflow (SEO), of size $\sim 0\farcm9$ with its axis aligned along $PA\sim45\arcdeg$, appears to emanate from the 
central source in IRAS\,05506.
\item The central source consists of a pseudo-disk and two compact outflow components, a a high-velocity outflow (HVO) and a medium-velocity outflow (MVO). 
The HVO (MVO), seen most clearly 
in the wings of the \codos J=2-1 (SO\,N,J=5,6-4,5 and SiO J=5-4) emission lines, is compact with an angular size 
$1\farcs4$ ($0\farcs9$), and has a (projected) outflow velocity of $\sim140$\,\kms~($\sim30$\,\kms).
\item Since the sense of the velocity gradient, the axis orientation, as well as the scalar momentum, of the HVO and SEO are roughly similar, 
the SEO is likely a result of the interaction of the HVO with the surrounding ambient cloud 
material.
\item The MVO emission produces prominent broad wing emssion in three of the four SO lines observed. 
The SO emission comes from a hot, shocked gas in the MVO at relatively high densities 
$\gtrsim10^7$\,cm$^{-3}$ and temperatures ($\gtrsim80$\,K). The lowest feasible value of the density is $\sim10^6$\,cm$^{-3}$, but 
then a high temperature ($>120$\,K) is needed.
\item The pseudo-disk has a size of about 9,600--11,100\,AU, and a thickness of 7,400\,AU. A simple disk model with infall and rotation that fits its observed 
position-velocity structure gives a 
central stellar mass of $8.3-18.8$\,\ms, and a disk mass of $0.64-0.85$\,\ms. The disk mass is consistent, within 
uncertainties, with that derived from the millimeter-wave thermal dust emission (1.7\,\ms). 
\item The wide-angle bullet spray shares a common axis with the pseudo-disk, and has an age comparable to 
that of MVO (few hunded years), suggesting that these three structures are intimately linked together.
\item A second-epoch imaging program with HST will be able to provide proper motions of nearby 
stars and  help in  
distinguishing between three possible models (dynamical decay of a stellar cluster, chance encounter of a runaway star with a dense cloud, and close passage of 
two protstars) to explain 
the IRAS\,05506 outflow source. 
\end{enumerate}

\acknowledgments
RS's contribution to the research described here was carried out at the Jet Propulsion Laboratory 
(JPL), California Institute of Technology, under a contract with NASA. Financial support was 
provided by NASA, in part from HST/GO awards (nos. GO-09463.01-A and GO-09801.01-A) from the Space 
Telescope Science Institute (operated by the Association of Universities for Research in Astronomy, 
under NASA contract NAS5-26555). CSC is supported by the Spanish MINECO through grants 
AYA2016-75066-C2-1-P and AYA2012-32032 and by the European Research Council through grant 
ERC-2013-SyG 610256. C.-F.L.  acknowledges grants from the Ministry of Science and Technology of 
Taiwan (MoST 104-2119-M-001-015-MY3) and the Academia Sinica (Career Development Award). The National Radio Astronomy Observatory 
is a facility of the National 
Science Foundation operated under cooperative agreement 
by Associated Universities, Inc..

\newpage
\begin{deluxetable}{lccccc}
\tablecolumns{6}
\tabletypesize{\normalsize}
\tablecaption{Interferometric Observing Log and Main Observational Parameters\label{tab:obs_log}}
\tablewidth{0pt}
\tablehead{
\colhead{Lines, Cont} & \colhead{Freq} & \colhead{Array} & \colhead{Beam\,(Maj$\times$Min)} & \colhead{PA}   & \colhead{Date} \\
\colhead{}            & \colhead{GHz}  &                 & \colhead{$({''})\times({''})$}   & \colhead{deg.} & \colhead{}    }
\startdata
SMA Observations & & & & & \\ 
\hline
CO 2-1\tablenotemark{a}          & 230.538 & Subcompact& $7\farcs05\,\times\,4\farcs11$& $71.49\arcdeg$  & 20-Mar-2010 \\ 
CO 2-1\tablenotemark{a}          & 230.538 & Extended  & $1\farcs34\,\times\,0\farcs81$& $81.46\arcdeg$  & 09-Feb-2012 \\ 
CO 2-1\tablenotemark{a}          & 230.538 & Ext+Sub   & $1\farcs62\,\times\,1\farcs01$& $72.97\arcdeg$  & ... \\
SiO 5-4 (v=0)   & 217.105 & Ext+Sub   & $1\farcs74\,\times\,1\farcs04$& $72.72\arcdeg$  & ... \\
SO 5,6-4,5      & 219.949 & Ext+Sub   & $1\farcs71\,\times\,1\farcs05$& $78.11\arcdeg$  & ... \\
Continuum       & 225     & Ext+Sub   & $1\farcs62\,\times\,1\farcs01$& $72.97\arcdeg$  & ... \\
\hline
OVRO Observations & & & & & \\
\hline
CO 1-0          & 115.271 & H+E+L   & $3\farcs12\,\times\,2\farcs68$& $-20.61\arcdeg$ & 31-Oct-2003 \\
Continuum       & 112     & H+E+L   & $2\farcs96\,\times\,2\farcs78$& $-33.37\arcdeg$ & 31-Oct-2003 \\
\enddata
\tablenotetext{a}{The $^{13}$CO (220.398\,GHz) and C$^{18}$O (219.560\,GHz) lines were also observed simultaneously; 
and the beam-sizes scale inversely with their frequencies.}
\end{deluxetable}

\clearpage
\begin{deluxetable}{lcccccc}
\tablecolumns{7}
\tabletypesize{\normalsize}
\tablecaption{Single-Dish Observing Log and Main Observational Parameters\label{tab:smt_obs_log}}
\tablewidth{0pt}
\tablehead{
\colhead{Lines} & \colhead{Freq} & \colhead{Tel.} & \colhead{T$_{sys}$} & $\eta _b$\tablenotemark{a}  &Backend & \colhead{Date} \\
\colhead{}      & \colhead{GHz}  &                & \colhead{K}         &            &        & \colhead{}     }
\startdata
CO 1-0          & 115.271 & ARO KP12m & 557  & 0.83 & 12M-MAC\tablenotemark{b} & 18-Nov-2007 \\ 
CO 2-1          & 230.538 & ARO SMT & 144  & 0.74 & CTS\tablenotemark{c}    & 18-Feb-2008 \\
CO 3-2          & 345.796 & ARO SMT & 1716 & 0.70 & CTS    & 20-Feb-2008 \\
CO 4-3          & 461.041 & CSO     & 2213 & 0.53 & FFTS\tablenotemark{d}    & 06-Mar-2013 \\
$^{13}$CO 1-0   & 110.201 & ARO KP12m & 207 & 0.83 & 12M-MAC & 20-Nov-2007 \\ 
$^{13}$CO 2-1   & 220.399 & ARO SMT & 163  & 0.74 & CTS    & 19-Feb-2008 \\
C$^{18}$O 1-0   & 109.782 & ARO KP12m & 222 & 0.83 & 12M-MAC & 21-Nov-2007  \\ 
SO 3,4-2,3      & 138.179 & ARO KP12m & 200 & 0.79   & 12M-MAC & 25-Nov-2007 \\
SO 4,3-3,2      & 158.972 & ARO KP12m & 205 & 0.76   & 12M-MAC & 28-Feb-2008 \\
SO 5,6-4,5      & 219.949 & ARO SMT & 150 & 0.74  & CTS    & 26-Jan-2008 \\
SO 6,7-5,6      & 261.844 & ARO SMT & 217 & 0.74   & CTS    & 19-Feb-2008 \\
SO$_2$ 4(3,1)-4(2,2)     & 221.965 & ARO SMT & 149 & 0.74 & CTS    & 18-Feb-2008 \\
SO$_2$ 12(1,12)-11(1,11) & 255.553 & ARO SMT & 181 & 0.74 & CTS    & 19-Feb-2008 \\
HCO$^{+}$ 1-0   & 89.1895 & ARO KP12m & 322 & 0.89 & 12M-MAC & 29-Nov-2007 \\
HCO$^{+}$ 3-2   & 267.558 & ARO SMT & 209 & 0.74  & CTS    & 19-Feb-2008 \\
CS 3-2          & 146.969 & ARO KP12m & 197 & 0.76  & 12M-MAC & 20-Nov-2007 \\
H$_2$CO 2(1,1)-1(1,0)    & 150.498 & ARO KP12m & 202 & 0.76 & 12M-MAC & 22-Jan-2008 \\
\enddata
\tablenotetext{a}{Beam-efficiencies for the ARO SMT and KP12m are taken from Tenenbaum et al. (2006, 2010), and Zhang et al. (2008). The CSO beam-efficiency 
was taken from http://www.submm.caltech.edu/cso/receivers/beams.html.}
\tablenotetext{b}{The Millimeter Autocorrelator (MAC) is a correlation spectrometer, and a number of different spectral resolutions are available. The 
intrinsic spectral resolution (bandwidth) was typically 24.4\,kHz (150\,MHz). Spectra were post-processed to a lower spectral spectral resolution  to increase 
the S/N ratio as appropriate.}
\tablenotetext{c}{The Chirp Transform Spectrometers (CTS) has a bandwidth of 215\,MHz, and a spectral resolution of 60--80\,kHz.  Spectra were post-processed 
to lower spectral spectral resolutions to increase the S/N ratio as appropriate.}
\tablenotetext{d}{The Fast Fourier Transform Spectrometer (FFTS), was used in its wide-band mode, which gave a bandwidth of 1000\,MHz, and a channel 
spacing of 122\,kHz. Spectra were post-processed to lower spectral spectral resolutions to increase the S/N ratio as appropriate.}
\end{deluxetable}

\begin{deluxetable}{lccccc}
\tablecolumns{6}
\tabletypesize{\scriptsize}
\tablecaption{Physical Structures associated with IRAS05506: Observed Properties\label{tab:phys_parm}}
\tablewidth{0pt}
\tablehead{
\colhead{Name} & \colhead{$V_{lsr}$ Range} & \colhead{Size}   & \colhead{$PA$}        & \colhead{Tracer} & \colhead{DataSet} \\
\colhead{}     & \colhead{(\kms)}   & \colhead{(${''}$)}      & \colhead{($\arcdeg$)} & \colhead{}       & \colhead{} }
\startdata
Central Star(s)  & ...              & ...                       & ... & ...            & ...        \\
Pseudo-disk      & $-6$ to 20      & $2\times2.8\tablenotemark{a}$ & 130\tablenotemark{b} & CO(2-1) & SMA Ext\\ 
MVO            & $-20$ to $-6$ \& 20 to 40     & $0.8\times1.0$  & ...                  & SO, SiO           & SMA Ext \\ 
HVO           & $-155$ to $-20$ \& 40 to 120  & 1.4              & 40\tablenotemark{b}  & CO(2-1)           & SMA Ext+Sub \\
Bullet Spray\tablenotemark{c}    & $-300$ to $100$      & 5.3             & $120-155$ &  H$\alpha$  & HST/WFC  \\
SEO & 0 to 5 \& 7 to 12          & 52                           & 45\tablenotemark{b}  & CO(2-1)           & SMT         \\
Cloud B        & 5 to 7          & 14   & 135\tablenotemark{d}      & $^{13}$CO(2-1),\,250\,\micron & SMA Ext+Sub, HSO \\
Cloudlet B1    & 7               & $>1.6,<15$      & ...         & CO J=3-2, 4-3  & SMT, CSO        \\
Cloud A core  & 5 to 7      & $\sim$125 & 45\tablenotemark{e}  & $^{13}$CO(2-1),\,250--500\,\micron  & SMT, HSO \\
Cloud A       & 5 to 7      & $>690(EW)\times310(NS)$  & $\sim90$\tablenotemark{c}  & $^{12}$CO(2-1)  & SMT         \\
" " "       & 5 to 7      & $>1800(EW)\times600(NS)$ & ...  & $^{12}$CO(1-0)  & KP12m      \\
\enddata
\tablenotetext{a}{(Vertical Thickness) $\times$ (Diameter)}
\tablenotetext{b}{$PA$ of symmetry axis}
\tablenotetext{c}{Data from Sahai et al. (2008)}
\tablenotetext{d}{$PA$ of long axis}
\tablenotetext{e}{$PA$ of hypotenuse, of this right-angled triangular-shaped structure}
\end{deluxetable}

\clearpage
\begin{deluxetable}{lcccc}
\tablecolumns{5}
\tabletypesize{\scriptsize}
\tablecaption{Physical Structures associated with IRAS05506: Derived Properties\label{tab:phys_prop}}
\tablewidth{0pt}
\tablehead{
\colhead{Name} & \colhead{Age}   & \colhead{Mass}   & \colhead{Scal.Mom.} &  \colhead{Energy} \\
\colhead{}     & \colhead{(yr)}  & \colhead{(\ms)}   & \colhead{(\momunit)}  & \colhead{($10^{45}$erg)} }
\startdata
Central Star(s)  & ...       &  $8.3-18.8$        & ...           & ...     \\
Pseudo-disk      & ...       &  $0.64-1.7$        & ...           & ...     \\ 
MVO              & 360       &  0.031             & 0.65          & 0.29    \\ 
HVO              & 130       &  0.074             & 9.8           & 7.8     \\
Bullet Spray\tablenotemark{a} & 230 & $4.6\times10^{-5}$ & 0.018   & 0.074   \\  
SEO              & 9300      &  1.6               & 4.7           & 0.3     \\
\enddata
\tablenotetext{a}{For knot $K1$, using data from Sahai et al. (2008), scaled to a distance D=3.7\,kpc}
\end{deluxetable}

\clearpage

\clearpage


\clearpage

\clearpage

\clearpage

\clearpage

\clearpage  

\clearpage

\clearpage

\clearpage



\clearpage

\clearpage

\clearpage

\clearpage

\clearpage
\begin{thebibliography}{}
\bibitem[Allen \& Burton(1993)]{1993Natur.363...54A} Allen, D.~A., \& Burton, M.~G.\ 1993, \nat, 363, 54
\bibitem[Bally et al. (2011)]{2011ApJ...727..113B} Bally, J., Cunningham, N.~J., Moeckel, N., et al.\ 2011, \apj, 727, 113
\bibitem[Bally et al.(2017)]{2017ApJ...837...60B} Bally, J., Ginsburg, A., Arce, H., et al.\ 2017, \apj, 837, 60 
\bibitem[Bally \& Zinnecker(2005)]{2005AJ....129.2281B} Bally, J., \& Zinnecker, H.\ 2005, \aj, 129, 2281
\bibitem[Bieging \& Peters(2011)]{2011ApJS..196...18B} Bieging, J.~H., \& Peters, W.~L.\ 2011, \apjs, 196, 18
\bibitem[Chatterjee \& Tan(2012)]{2012ApJ...754..152C} Chatterjee, S., \& Tan, J.~C.\ 2012, \apj, 754, 152
\bibitem[Goddi et al.(2011)]{2011ApJ...728...15G} Goddi, C., Humphreys, E.~M.~L., Greenhill, L.~J., Chandler, C.~J., \& Matthews, L.~D.\ 2011, \apj, 728, 15
\bibitem[Goldsmith et al.(1997)]{1997ApJ...491..615G} Goldsmith, P.~F., Bergin, E.~A., \& Lis, D.~C.\ 1997, \apj, 491, 615
\bibitem[Hartigan et al.(1987)]{hart87} Hartigan, P., Raymond, J., \& Hartmann, L.\ 1987, ApJ, 316, 323
\bibitem[Hogerheijde(2001)]{2001ApJ...553..618H} Hogerheijde, M.~R.\ 2001, \apj, 553, 618
\bibitem[Lee et al. (2014)]{lee14} Lee, C.-F., Hirano, N., Zhang, Q., et al.\ 2014, \apj, 786, 114
\bibitem[Lee et al.(2006)]{2006ApJ...639..292L} Lee, C.-F., Ho, P.~T.~P., Beuther, H., et al.\ 2006, \apj, 639, 292
\bibitem[Lonsdale et al.(1982)]{1982AJ.....87.1819L} Lonsdale, C.~J., Becklin, E.~E., Lee, T.~J., \& Stewart, J.~M.\ 1982, \aj, 87, 1819 
\bibitem[Luhman et al.(2017)]{2017ApJ...838L...3L} Luhman, K.~L., Robberto, M., Tan, J.~C., et al.\ 2017, \apjl, 838, L3 
\bibitem[Lumsden et al.(2013)]{2013ApJS..208...11L} Lumsden, S.~L., Hoare, M.~G., Urquhart, J.~S., et al.\ 2013, \apjs, 208, 11
\bibitem[Molinari et al.(2016)]{2016A&A...591A.149M} Molinari, S., Schisano, E., Elia, D., et al.\ 2016, \aap, 591, A149 
\bibitem[Plambeck et al.(2009)]{2009ApJ...704L..25P} Plambeck, R.~L., Wright, M.~C.~H., Friedel, D.~N., et al.\ 2009, \apjl, 704, L25
\bibitem[Plambeck \& Wright(2016)]{2016ApJ...833..219P} Plambeck, R.~L., \& Wright, M.~C.~H.\ 2016, \apj, 833, 219
\bibitem[Podio et al.(2015)]{2015A&A...581A..85P} Podio, L., Codella, C., Gueth, F., et al.\ 2015, \aap, 581, A85
\bibitem[Qiu et al. (2008)]{qiu08} Qiu, K. et al. 2008, ApJ 685, 1005
\bibitem[Sahai et al.(2008)]{sah08} Sahai, R., Claussen, M., S{\'a}nchez Contreras, C., Morris, M., \& Sarkar, G.\ 2008, \apj, 680, 483
\bibitem[Sahai et al.(1999)]{1999ApJ...514L.115S} Sahai, R., te Lintel Hekkert, P., Morris, M., Zijlstra, A., \& Likkel, L.\ 1999, \apjl, 514, L115
\bibitem[Sridharan et al.(2002)]{2002ApJ...566..931S} Sridharan, T.~K., Beuther, H., Schilke, P., Menten, K.~M., \& Wyrowski, F.\ 2002, \apj, 566, 931
\bibitem[S{\'a}nchez Contreras et al.(2008)]{2008ApJS..179..166S} S{\'a}nchez Contreras, C., Sahai, R., Gil de Paz, A., \& Goodrich, R.\ 2008, \apjs, 179, 
166-194 
\bibitem[Rodr{\'{\i}}guez et al.(2005)]{2005ApJ...627L..65R} Rodr{\'{\i}}guez, L.~F., Poveda, A., Lizano, S., \& Allen, C.\ 2005, \apjl, 627, L65
\bibitem[S{\'a}nchez Contreras \& Sahai(2012)]{2012ApJS..203...16S} S{\'a}nchez Contreras, C., \& Sahai, R.\ 2012, \apjs, 203, 16
\bibitem[Tan(2004)]{2004ApJ...607L..47T} Tan, J.~C.\ 2004, \apjl, 607, L47
\bibitem[Tenenbaum et al.(2006)]{2006ApJ...649L..17T} Tenenbaum, E.~D., Apponi, A.~J., Ziurys, L.~M., et al.\ 2006, \apjl, 649, L17 
\bibitem[Tenenbaum et al.(2010)]{2010ApJS..190..348T} Tenenbaum, E.~D., Dodd, J.~L., Milam, S.~N., Woolf, N.~J., \& Ziurys, L.~M.\ 2010, \apjs, 190, 348 
\bibitem[van der Tak et al.(2007)]{2007A&A...468..627V} van der Tak, F.~F.~S., Black, J.~H., Sch{\"o}ier, F.~L., Jansen, D.~J., \& van Dishoeck, E.~F.\ 2007, 
\aap, 468, 627 
\bibitem[Wilson (1999)]{wils99} Wilson, T. L. 1999, Rep. Prog. Phys., 62, 143
\bibitem[Zapata et al. (2009)]{zapata09} Zapata, L.~A., Schmid-Burgk, J., Ho, P.~T.~P., Rodr{\'{\i}}guez, L.~F., \& Menten, K.~M.\ 2009, \apjl, 704, L45
\bibitem[Zapata et al.(2012)]{2012ApJ...754L..17Z} Zapata, L.~A., Rodr{\'{\i}}guez, L.~F., Schmid-Burgk, J., et al.\ 2012, \apjl, 754, L17 
\bibitem[Zapata et al.(2013)]{2013ApJ...765L..29Z} Zapata, L.~A., Schmid-Burgk, J., P{\'e}rez-Goytia, N., et al.\ 2013, \apjl, 765, L29 
\bibitem[Zhang et al.(2008)]{2008ApJ...678..328Z} Zhang, Y., Kwok, S., \& Dinh-V-Trung 2008, \apj, 678, 328-346 
\bibitem[Zinnecker \& Yorke(2007)]{2007ARA&A..45..481Z} Zinnecker, H., \& Yorke, H.~W.\ 2007, \araa, 45, 481 
\end{thebibliography}
\end{document}